\documentclass[onecolumn]{aastex6}

\usepackage{hyperref}

\usepackage[utf8x]{inputenc}
\usepackage[english]{babel}
\usepackage{fontenc}
\usepackage{graphicx}

\usepackage{natbib}
\usepackage{amsmath}
\usepackage{amsfonts}
\usepackage{amssymb}
\usepackage{mathrsfs}

\usepackage{wasysym}  

\usepackage{siunitx}

\begin{document}

\title{Accretion of Saturn's inner mid-sized moons from a massive primordial ice ring}

\author{J.~Salmon and R.~M.~Canup}
\affil{Southwest Research Institute\\Planetary Science Directorate\\1050 Walnut Street, Suite 300, Boulder, CO 80302, USA}
\email{julien@boulder.swri.edu}

\begin{abstract}
Saturn's rings are rock-poor, containing 90 to 95\% ice by mass. As a group, Saturn's moons interior to and including Tethys are also about 90\% ice. Tethys itself contains $< 6\%$ rock by mass, in contrast to its similar-mass outer neighbor Dione, which contains $> 40\%$ rock. Here we simulate the evolution of a massive primordial ice-rich ring and the production of satellites as ring material spreads beyond the Roche limit. We describe the Roche-interior ring with an analytic model, and use an N-body code to describe material beyond the Roche limit. We track the accretion and interactions of spawned satellites, including tidal interaction with the planet, assuming a tidal dissipation factor for Saturn of $Q \sim 10^4$. We find that ring torques and capture of moons into mutual resonances produces a system of ice-rich inner moons that extends outward to approximately Tethys's orbit in $10^9$ years, even with relatively slow orbital expansion due to tides. The resulting mass and semi-major axis distribution of spawned moons resembles that of Mimas, Enceladus and Tethys. We estimate the mass of rock delivered to the moons by external cometary impactors during a late-heavy bombardment. We find that the inner moons receive a mass in rock comparable to their current total rock content, while Dione and Rhea receive an order-of-magnitude less rock than their current rock content. This suggests that external contamination may have been the primary source of rock in the inner moons, and that Dione and Rhea formed from much more rock-rich source material. Reproducing the distribution of rock among the current inner moons is challenging, and appears to require large impactors and stochasticity and/or the presence of some rock in the initial ring.    
\end{abstract}

\maketitle

\section{Introduction}
Saturn's satellites display a diversity of masses and compositions that is challenging to explain. Massive Titan likely formed in a primordial subnebula surrounding Saturn as the planet completed its gas accretion \citep[e.g.][]{canup06}. The small icy moons orbiting close to the rings, from Atlas to Janus, appear to have formed relatively recently from ring material that collisionally spread beyond the Roche limit \citep{charnoz10}. The origin of the mid-sized moons exterior to Janus and interior to Titan -- including Mimas, Enceladus, Tethys, Dione and Rhea -- is less clear. Some or all of them may have accreted directly from Saturn's subnebula.  Such an origin would generally imply compositions that are roughly half rock and half ice, reflecting the expected solar composition of material inflowing to the subnebula.  Instead the mid-sized moons have a broad range of densities, with Mimas and Tethys being extremely ice-rich with little or no rock (Table 1). The rings are continually contaminated by micrometeoroid bombardment, which has increased their rock content over time \citep{cuzzi98}. That they remain so ice-rich even after this contamination implies that the rings were essentially pure ice when they formed.  

The mass of Saturn's current rings is about a few $\times 10^{19}$ to perhaps $10^{20}$ kg \citep{robbins10}. Traditional models for the origin of Saturn's rings envisioned an initial ring mass comparable to that of the current rings, and invoke either the collisional disruption of a small, Mimas-sized moon orbiting within the Roche limit by an external impactor \citep{harris84,charnoz09}, or the tidal disruption of a cometary interloper that passed very close to Saturn \citep{dones91}. However it is now appreciated that Saturn's rings could have initially been much more massive. Local gravitational instabilities within a massive ring produce a viscosity that is proportional to the square of the ring's surface density \citep{ward78,daisaka01}, so that a massive ring spreads rapidly at first but then slows as its surface density decreases.  Simulations show that as a massive ring at Saturn viscously spreads, its mass asymptotically approaches that of the current rings over 4.5 Gyr \citep{salmon10}, with the overwhelming majority of the ring's initial mass either accreted by Saturn or driven outward beyond the Roche limit. The latter would provide a natural source of material to ``spawn'' moons from the outer edge of the rings.   

To spawn moons as massive as Tethys, Dione, or Rhea implies an initial ring containing $\sim 10^{21}$ to $10^{22}$ kg, some 10 to $10^2$ times more massive than the current rings \citep{canup10,charnoz11}. Collisional disruption of a Roche-interior moon would be very unlikely to produce such a massive ring, because a massive moon would remain within the Roche limit for only a short time due to its rapid tidal evolution, e.g., for only a few million years for a $10^{22}$ kg satellite and slow tidal evolution.  A disruptive collision by an external impactor during such a brief period would be extremely improbable.  A massive ring could be produced by tidal disruption during the close passage of a Titan-sized comet by Saturn \citep{hyodo16}.  However the background population of extremely large comets needed to make such an event probable at Saturn would also imply that similar encounters at Uranus and Jupiter should have produced massive ring systems around those planets too, and no such massive ring systems exist there today.  

Alternatively a massive ring at Saturn can be produced by tidal stripping from a large primordial satellite \citep{canup10}.  Models of satellite accretion within the Saturnian subnebular disk suggest that Titan-sized satellites spiraled into Saturn due to density wave interactions with the gas component of the disk, i.e., through Type I migration \citep{canup06,sasaki10,ogihara12}. As it spiraled towards the planet, a Titan-sized moon would most likely have a differentiated interior, with an ice mantle overlying a rocky core, due to the energy of its accretion and strong tidal heating \citep{canup10}. Tidal mass loss would begin once the satellite migrated within the Roche limit set by its mean density, located at $\approx 1.75R_{\saturn}$ for a satellite composed of roughly half rock, half ice, where $R_{\saturn} = \SI{58232}{km}$ is Saturn's current mean radius. Tides would initially strip material from the satellite's outer ice shell.  The removal of low-density ice would cause the satellite's mean density to increase until the remnant satellite became marginally stable at a given orbital distance \citep{canup10}.  Continued inward migration would then lead to additional ice removal.  Tidal stripping would continue until either the remnant satellite collided with the planet, or its higher-density rocky core disrupted as the satellite passed within the Roche limit for rock, depending on which event occurred first.   

The Roche limit for rock of density $\rho_{rock}$ is at $a_{R,rock} = 1.5R_{\saturn}(\SI{3}{\gram\per\cubic\centi\metre}/\rho_{rock})^{1/3}$.  Planet contraction models suggest that Saturn's radius at the time of the dispersal of the solar nebula would have been between $R_p = 1.5R_{\saturn}$ and $R_p = 1.7R_{\saturn}$ \citep[e.g.][see also Fig. \ref{figSynchronousOrbit}]{marley07,fortney07}.  For $a_{R,rock} \le R_p$, the remnant satellite would collide with the planet before its rocky core disrupts, and in this case tidal stripping produces an essentially pure ice ring, in agreement with the unusually ice-rich composition of the rings today \citep{canup10}. The mass of the ice ring so produced depends on the relative position of the planet's surface compared to the Roche limit for rock.  In the limiting case that $a_{R,rock} = R_p$, tidal stripping from a Titan-sized satellite produces an ice ring with $\sim 10^{22}$ kg \citep{canup10}; a less massive ice ring results if $a_{R,rock} < R_p$.   

Rings produced by tidal stripping while a planet is still accreting substantial gas through its circumplanetary disk would likely be lost due to gas drag.  This may have been the fate of massive rings produced at Jupiter from satellites that spiraled into the planet before the Galilean moons formed.  While Jupiter has massive inner moons that survived (Io and Europa), Saturn does not.  The lack of an inner Titan-sized moon at Saturn would be expected if large inner moons spiraled into Saturn as gas accretion by the planet was ending \citep{canup06}.  A massive ring produced at the end of gas accretion can survive against gas drag because its surface density is orders-of-magnitude larger than that of the dispersing gas disk \citep{canup10}.  Thus tidal stripping is consistent with the production of a long-lived massive ring at Saturn, while similarly-produced structures at Jupiter (and Uranus, if it too accreted gas through a disk) could well have been lost.

As a massive ring viscously spreads, material driven beyond the Roche limit can accrete into satellites.  The mass and orbital distribution of satellites spawned from a ring depend on the ring surface density and the rate of tidal dissipation in Saturn, because the latter controls the rate of satellite orbital expansion due to tides raised on Saturn.  If Mimas tidally expanded to its current distance over 4.5 Gyr, a time-average tidal parameter for Saturn of $Q > 1.8 \times 10^4$ is implied \citep{murray99}.  For $Q \sim 10^4$ to $10^5$, initial estimates suggested that a $\sim 10^{22}$ kg ring could spawn analogs to Mimas, Enceladus and Tethys \citep{canup10}.  Subsequent detailed simulations considered more rapid tidal evolution with $Q \sim 10^3$, and found that in this case the masses and positions of all of the mid-sized moons, including outermost Rhea and Dione, could be explained as byproducts of a massive ring's expansion \citep{charnoz11}. Such a low value for Saturn's $Q$ has been inferred from astrometric observations of its satellites over the last $10^2$ yr \citep{lainey12,lainey15}, but it remains challenging to explain and its applicability to primoridal Saturn is unclear, because Saturn's $Q$ may have varied by orders-of-magnitude over the age of the Solar System \citep{Wu05,fuller16}.

In addition to the masses and orbital spacings of the mid-sized moons, any origin model must also account for their varied densities and compositions, which do not follow simple trends with either satellite mass or orbital distance. Nonetheless we argue that it is useful to consider two groupings based on the total mass of rock in each object (Table \ref{table_moons_properties}). The inner three moons (Mimas, Enceladus and Tethys) each contain $\le \SI{6e19}{\kilogram}$ in rock. Mimas and Tethys are overwhelmingly icy.  While Enceladus is currently proportionally rock-rich, it may have lost substantial ice (perhaps comparable to its present mass) over its history if its current thermal activity has been typical.  Thus Enceladus could have been more ice-rich when it formed.  In contrast the outer two moons, Dione and Rhea, contain an order-of-magnitude more rock, $\sim 5$ to $8 \times 10^{20}$ kg each. The distinction between these two groupings is particularly notable when comparing neighboring Tethys and Dione. {\it Despite differing in mass by less than a factor of two, Tethys contains essentially no rock while Dione is roughly half rock}. Either Tethys and Dione formed through a similar process but somehow acquired overwhelmingly different rock masses, or they represent different formation processes. The former was advocated in \cite{charnoz11}; we pursue the latter possibility here. 

\begin{deluxetable}{l c c c c c }
	\tabletypesize{\footnotesize}
	\tablewidth{0pt}
	\tablecolumns{6}
	\tablecaption{Some properties of  Saturn's mid-size moons. \label{table_moons_properties}}
	\tablehead{ & \colhead{Distance} & \colhead{Mass} & \colhead{Density} & \colhead{Rock mass fraction} & \colhead{Mass of rock} \\
		& \colhead{$\left(R_{\saturn}\right)$} & \colhead{$\left(\SI{e-6}{M_{\saturn}}\right)$} & \colhead{$\left(\si{g.cm^{-3}}\right)$} & \colhead{$(\%)$} & \colhead{$\left(\SI{e19}{\kilogram}\right)$}}
	\startdata
	Mimas & 3.18 & \num{0.066} & 1.15 & $17 - 29$ & $0.64 - 1.1$\\
	Enceladus & 4.09 & \num{0.19} & 1.61 & $52 - 61$ & $5.6 - 6.6$\\
	Tethys & 5.06 & \num{1.09} & 0.97 & $0 - 6$ & $0 - 3.7$\\
	Dione & 6.47 & \num{1.93} & 1.48 & $42 - 52$ & $46 - 57$\\
	Rhea  & 9.05 & \num{4.06} & 1.23 & $25 - 35$ & $58 - 81$\\
	\enddata
	\tablecomments{Distance, mass, density, estimated mass of rock and rock mass fraction of Saturn's mid-size moons. $M_{\saturn}=\SI{568.46e24}{\kilogram}$ and $R_{\saturn}=\SI{58232}{\km}$ are Saturn's mass and mean radius. Mimas, Enceladus and Tethys all have $\le \SI{7e19}{\kilogram}$ in rock, while the outer two (Dione and Rhea) each have about an order-of-magnitude more. Enceladus's density may have been lower in the past if it has lost substantial ice through geophysical activity.}
\end{deluxetable}

In this paper, we simulate the viscous evolution of a massive ice ring and the accompanying accretion and tidal evolution of satellites spawned from its outer edge, assuming $Q \ge 10^4$.  We consider a ring that is essentially pure ice\footnote{Tidal stripping from a completely differentiated ice-rock satellite can produce a pure ice ring.  However the ice mantle of an incompletely differentiated satellite could contain a component of rock.  Rock fragments descending via Stokes flow have a settling rate proportional to the square of the fragment radius. Thus large chunks are rapidly lost, while small (less than kilometer-sized for a Titan-like satellite; \cite{barr08}) rocky fragments could plausibly be embedded within the ice tidally stripped from a satellite's outer layers.  The mass fraction of such fragments in the initial ring would be limited to less than a few to ten percent, based on the current rock content of Saturn's rings.}, which would spawn predominantly icy satellites.   We assume that Rhea and Dione formed separately, e.g., as direct accretional products from the Saturnian subnebula \citep[e.g., as in][]{canup06}.   We first determine whether a massive ice ring can produce good analogs to Mimas, Enceladus and Tethys in terms of satellite mass and orbital radius. We then estimate the delivery of rock to the mid-sized moons by external impactors during a late heavy bombardment (LHB) to assess whether this process could supply the inner moon's rock component, as suggested in \citep{canup13}.

Overall we explore a similar problem as in \cite{charnoz11}, with key differences.   We consider slow tidal evolution and an initial ice ring, while they considered rapid tidal evolution ($Q\sim 10^3$) and an initial ring that contains large, $\sim 10^2$-km chunks of rock comprising a substantial portion of its total mass. We postulate that the inner three mid-sized moons (or their progenitors) were spawned from the rings, while \cite{charnoz11} propose that all of the mid-sized moons out to and including Rhea originated in this manner.   The simulation methods are also different, and complementary.  The \cite{charnoz11} model describes the ring's evolution with a 1D Eulerian hydrodynamical model that evolves the ring's radial surface density profile due to viscosity and resonant torques with exterior moons.  Their companion model for the growth of moons is simple, an analytic treatment that does not explicitly treat moon-moon interactions.  In our simulations, the ring model is simple and analytic, assuming a uniform surface density ring whose total mass and outer edge position evolve with time due to viscosity and resonant torques \citep{salmon12}.  While we include all the same resonances as in the Charnoz model, our calculation of the resonant torque is less accurate because in reality the ring's surface density would vary with orbital radius.  Instead we focus computational effort on the accretion process, which we describe by a full $N-$body model.  This allows us to directly simulate the capture of moons into mutual mean motion resonances and the accompanying growth in satellite eccentricities as satellites are tidally driven outward, which ultimately will affect the stability of spawned satellite systems \citep{peale15}.  We also consider the early temporal evolution of Saturn's radius and synchronous orbit, while \cite{charnoz11} assume Saturn's current radius and synchronous orbit location.

In Section 2 we describe our numerical model. In Section 3 we present results of our simulations for various initial conditions, tracking the system's evolution for $10^8$ yr.  We explore the influence of Dione and Rhea on the accretion and evolution of the inner mid-sized moons, as well as the inclusion of tidal dissipation within the growing moons.  Each $10^8$ yr simulation involves integration of about $7 \times 10^{10}$ orbits at the Roche limit, requiring months of CPU time.  In Section 4, we present follow-on integrations that consider an accelerated evolution to approximate the behavior of the resulting ring-satellite systems over $10^9$ yr. In Section 5 we estimate the delivery of rock to the mid-sized moons during an LHB \citep{gomes05}, and in Section 6 we discuss the overall findings.

\section{Numerical model}
\subsection{Coupled ring-satellite accretion simulation}
The core numerical model used here is based on one developed to study the accretion of the Earth's Moon from a protolunar disk \citep{salmon12,salmon14}; additional details are contained in \cite{salmon12} and Appendices therein. The code couples an analytical model of a viscous interior ring to the $N$-body code SyMBA \citep{duncan98}, which is used to simulate the accretion of moons exterior to the ring.  The inner ring extends from the planet's surface at radius $R_p$ to an outer edge $r_{out}$, which is initially set equal to the Roche limit, $a_R=1.524\left(M_P/\rho\right)^{1/3}$, where $M_P$ is the planet's mass and $\rho$ is the density of ring material.  We consider ice ring particles with $\rho = \SI{0.9}{\gram\per\cubic\centi\metre}$ so that $a_R = 2.24R_{\saturn}$. The ring's surface density, $\sigma$, is assumed to be constant across the radial extent of the ring, with
\begin{equation}
\sigma = \frac{M_r}{\pi (r_{out}^2-R_p^2)},
\end{equation}  
where $M_r$ is the ring's total mass.  We emphasize the distinction between Saturn's early radius ($R_p$) and its current mean radius ($R_{\saturn}$), where in our simulations $R_p$ is larger than $R_{\saturn}$ because we consider a primordial ring and a young Saturn.  An initial ice ring with $M_r = 10^{22}$ kg, $r_{out} = a_R$, and $R_p = 1.4R_{\saturn}$ has $\sigma \approx \SI{3e4}{\gram\per\square\centi\metre}$.   With time, $M_r$, $\sigma$, and $r_{out}$ vary due to the ring's viscosity and interactions with outer moons.  

The ring spreads with a viscosity $\nu$ that includes the effects of self-gravity \citep{ward78,salo95,daisaka99,daisaka01}:
\begin{equation}
\nu \approx \frac{\pi^2G^2\sigma^2}{\Omega^3},
\end{equation}
where $G=\SI{6.67e-11}{\cubic\metre\per\kilo\gram\per\square\second}$is the gravitational constant and $\Omega = \sqrt{\left(GM_P/r^3\right)}$ is the orbital frequency at distance $r$ from the planet's center. In our calculation of $\Omega$ for the viscosity, we set $r=r_{out}$.  Near the ring's inner edge, the viscosity would be lower for a fixed surface density because $r$ is smaller. More detailed models of the rings' viscous evolution predict formation of a inner density peak \citep{salmon10}, which would tend to increase the viscosity through a higher value of $\sigma$. Our model does not resolve the radial structure of the rings, and applies the same viscosity across the entirety of the ring.  Viscous spreading causes the ring to lose mass through its inner edge as mass is accreted by the planet, and causes the outer edge of the ring to expand \citep[see Appendix A in][for details]{salmon12}.

The $N$-body portion of the code tracks the orbital and collisional evolution of discrete objects beyond the Roche limit. Each outer object interacts with the inner ring at its strongest Lindblad resonances, resulting in a positive torque on the object and a negative torque on the ring, which causes $r_{out}$ to contract. The total torque $T_{res}$ exerted by the rings on an exterior satellite per unit satellite mass is found by summing the torques due to all the $0^{th}$ order resonances that fall in the disk \citep{salmon12}:

\begin{equation}
\frac{T_{res}}{m} = \left(\frac{\pi^2}{3}\mu G\sigma a\right)C(p),
\label{equ_disktorque_on_sat}
\end{equation}
where $m$ is the satellite's mass, $\mu = m/M_{\saturn}$, $C(p) = \sum_{p=2}^{p_*}2.55p^2(1-1/p)$ and $p_*$ is the highest $p$ for which resonance $(p:p-1)$ falls in the disk. Once an object is far enough from the ring that its strongest resonances no longer fall within the ring, which occurs for orbital radii $\ge 1.6r_{out}$, it no longer interacts directly with the ring. The net change in the position of the ring's outer edge at each time step is found by considering the combined effect of resonant torques due to all of the exterior objects and the ring's viscosity.  If the former dominates, the ring edge contracts, while if the latter dominates, $r_{out}$ expands.  

Ring material that spreads beyond the Roche limit can clump into tidally stable fragments due to local gravitational instabilities, which then mutually collide and accrete into still larger objects.  The mass $m_f$ of a fragment formed via local instability is \citep{goldreich73}:
\begin{equation}
m_f \approx \frac{16\pi^4\xi^2\sigma^3r_{out}^6}{M_p^2} \approx \SI{5.8e13}{\gram}\left(\frac{r_{out}}{a_R}\right )^6
\left( \frac{\sigma}{\SI{3e4}{\gram\per\square\centi\metre}}\right)^3 ,
\end{equation}
 where $\xi$ is a factor of order, but less than, unity. This mass is of order $10^{-13}$ times Saturn's mass, which is too small to be feasibly treated in our $N-$body simulations.  Small fragments would likely collide and merge rapidly into bigger objects \citep{charnoz10}. In our simulations, we set the mass of objects spawned at the Roche limit to $m_f = \SI{e-8}{M_{\saturn}}$, which is about one tenth Mimas' mass, so that the growth of Mimas-sized moons is (marginally) resolved. When $r_{out} > a_R$, we remove a mass $m_f $ from the ring, and add a new discrete object having this mass to the $N$-body code at $r \approx r_{out}$.  The position of the ring's outer edge is then decreased to conserve angular momentum, such that $L_d + L_f = L_{d,0}$, where $L_{d,0}$ and $L_d$ are the angular momentum of the ring before and after the formation of the fragment, respectively, and $L_f$ is the orbital angular momentum of the newly formed object.

We use tidal accretion criteria \citep{ohtsuki93,canup95} to determine if collisions between objects in the $N$-body code will result in a merger or intact rebound, assuming completely inelastic collisions. The outcome of a collision then depends on the impact energy, the mass ratio of the colliding bodies, and the orbital distance of the impact with respect to the Roche limit.  We use a ``total accretion'' criterion, in which we assume that collisions occur in the radial direction along the widest axis of the Hill sphere of the colliding bodies, which is the most favorable case for accretion.

It is possible that interactions among orbiting bodies can cause an object to be scattered onto an orbit whose pericenter is close to the planet.  In the limiting case of an inviscid fluid object on a parabolic orbit, tidal disruption will occur in a single pass once its pericenter $r_p$ satisfies \citep{sridhar92}:
\begin{equation}
r_p < 1.05 \left(\frac{M_p}{\rho}\right)^{1/3} \approx 0.7a_R.
\end{equation}
When an object satisfies this criterion, we remove it from the $N$-body code and add its mass and angular momentum to the ring.  In practice, such events occur rarely in our simulations.

\subsection{Tidal evolution}
Because the evolution of the ring and the associated growth of spawned satellites occurs over $\ge 10^8$ yr, evolution of the satellite orbits due to tidal interaction with Saturn must be considered.  Tides raised on Saturn by a satellilte orbiting exterior (interior) to synchronous orbit produce a positive (negative) torque on the satellite's orbit, causing its orbit to expand (contract).  The current synchronous orbit -- where the orbital period equals Saturn's rotational day -- lies within the Roche limit, with $a_{sync} = 1.9R_{\saturn}$.  However early Saturn's radius was larger and by conservation of angular momentum it would have been rotating more slowly, with $a_{sync}$ outside the Roche limit \citep{canup10}. Tides raised on a satellite by the planet also modify the satellite's orbit, predominantly acting to decrease its orbital eccentricity.      

We include the modification of satellite orbits due to tidal evolution in the $N$-body portion of our code by applying an additional accelerating ``kick'' to each orbiting object at every time step \citep{canup99}. We utilize the constant time delay tidal model of Mignard, which has a relatively straightforward analytic form, and is valid for orbits near or that cross $a_{sync}$, and for high orbital eccentricities.  

\subsubsection{Planetary tides}
The acceleration of a satellite of mass $m$ due to the second-order distortion it raises on the planet is given by \citep{mignard80,touma94}:

\begin{equation}
\left.\frac{d^2\mathbf{r}}{dt^2}\right|_p=-\frac{3k_2GmR_p^5}{r^{10}}\left(1+\frac{m}{M_p}\right)\Delta t\left[2\left(\mathbf{r}\cdot\mathbf{v}\right)\mathbf{r}+r^2\left(\mathbf{r}\times\boldsymbol{\omega}+\mathbf{v}\right)
\right],
\label{equTidalAcc}
\end{equation}
where $\mathbf{r}=\left(x,y,z\right)$ and $\mathbf{v}=\left(v_x,v_y,v_z\right)$ are the planetocentric satellite's position and velocity, $k_2$ is the planet's second order Love number, and $\boldsymbol{\omega}=\omega \bf{u_z}$ is the planet's spin vector that we assume lies along the z-axis. The early value of $k_2$ is unknown; we adopt its current value for Saturn, $k_2 = 0.32$. The time lag $\Delta t$ is defined as the time between the tide raising potential and when the equilibrium figure is achieved in response to this potential. The relation between the tidal time lag and the tidal dissipation factor $Q$ is $Q \sim \left(\psi \Delta t\right)^{-1}$ for a system oscillating at frequency $\psi$.  For the planet, the dominant frequency is $\psi = 2\left|\omega-n\right|$, where $n$ is the satellite's mean motion, with $\Delta t \sim 1/(2\left|\omega-n\right| Q)$.

\subsubsection{Satellite tides}
The acceleration on a satellite due to tides raised by the planet on the satellite is \citep{mignard80}:

\begin{equation}
\left.\frac{d^2\mathbf{r}}{dt^2}\right|_s=-\frac{3k_2GmR_p^5}{r^{10}}\left(1+\frac{m}{M_p}\right)\mathcal{A}\Delta t\left[2\left(\mathbf{r}\cdot\mathbf{v}\right)\mathbf{r}+r^2\left(\mathbf{r}\times\boldsymbol{\omega_s}+\mathbf{v}\right)\right],
\end{equation}
where $\boldsymbol{\omega_s}$ is the satellite's spin vector. The factor $\mathcal{A}$ reflects the strength of satellite versus planetary tides, with
\begin{equation}
	\mathcal{A}=\left(\frac{m}{M_p}\right)^{-2} \left(\frac{R_s}{R_p}\right)^5 \left(\frac{k_{2s}}{k_2}\right) \left(\frac{\Delta t_s}{\Delta t}\right),
\end{equation}
where $k_{2s}$, $\Delta t_s$ and $R_s$ are the satellite's Love number, tidal time lag and physical radius. 

The appropriate value for $\mathcal{A}$ is very uncertain.  In our simulations, $10 \le (M_p/m)^2(R_s/R_p)^5 \le 10^2$.  Estimates suggest $10^{-3} \le k_{2s} \le 10^{-1}$ for icy satellites \citep[][Table 4.1]{murray99}.  For satellite tides, $\psi \approx n$ and $\Delta t_s \sim 1/(Q_sn)$, where we consider a satellite tidal dissipation factor $Q_s \sim 10^2$.   For $\left|\omega/n-1\right| \sim 10^{-1}$, the final term in the expression above for $\mathcal{A}$ is of order $10 \le (\Delta t_s/\Delta t) \le 10^2$ for $10^4 \le Q \le 10^5$.  Thus the plausible range for $\mathcal{A}$ is of order $10^{-1} \le \mathcal{A} \le 10^3$. As such we perform two sets of simulations that consider limiting cases: one without satellites tides ($\mathcal{A} = 0$), and one with strong satellite tides ($\mathcal{A} =1000$).

When computing $d^2\mathbf{r}/dt^2|_s$, we make the simplifying assumption that satellites are rotating synchronously, so that $\omega_s \approx n = \sqrt{GM_p/a^3}$, where $a$ is semi-major axis.  A non-synchronously rotating, uniform density satellite on a circular, non-inclined orbit will experience a torque $N=C\dot{\omega}_s$, where $C =(2/5)mR_s^2$ is the satellite's moment of inertia and $\dot{\omega}_s$ is the time rate of change of its rotation rate, given by \citep{peale15}:
\begin{equation}
	\frac{d\omega_s}{dt}=-\frac{3k_{2s}GM_p^2R_s^5}{Ca^6}\Delta t_s(\omega_s -n)=-\frac{15}{2}k_{2s}\frac{M_p}{m}\frac{R_s^3}{a^3}\frac{n}{Q_s}(\omega_s-n).
\end{equation}
If $n$ is nearly constant, the quantity $(\omega_s - n)$ decays exponentially with a time constant
\begin{equation}
	\tau_{despin} = \frac{2}{15}\frac{1}{k_{2s}}\frac{m}{M_p}\frac{a^3}{R_s^3}\frac{Q_s}{n}
	\approx 2 \left(\frac{Q_s/k_{2s}}{10^3}\right)\left(\frac{m/M_p}{10^{-7}}\right)\left(\frac{a}{3R_{\saturn}}\right)^{9/2}\left(\frac{250^{} {\rm - km}}{R_s}\right)^3 {\rm years}
\end{equation}
Thus $\omega_s$ will likely approach a synchronous value on a timescale short compared to orbital migration timescales. 

\subsection{Saturn's early radius and synchronous orbit location} \label{Subsection_SaturnRadius}

\begin{figure}[!h]
\begin{center}
\includegraphics[width=8.5cm]{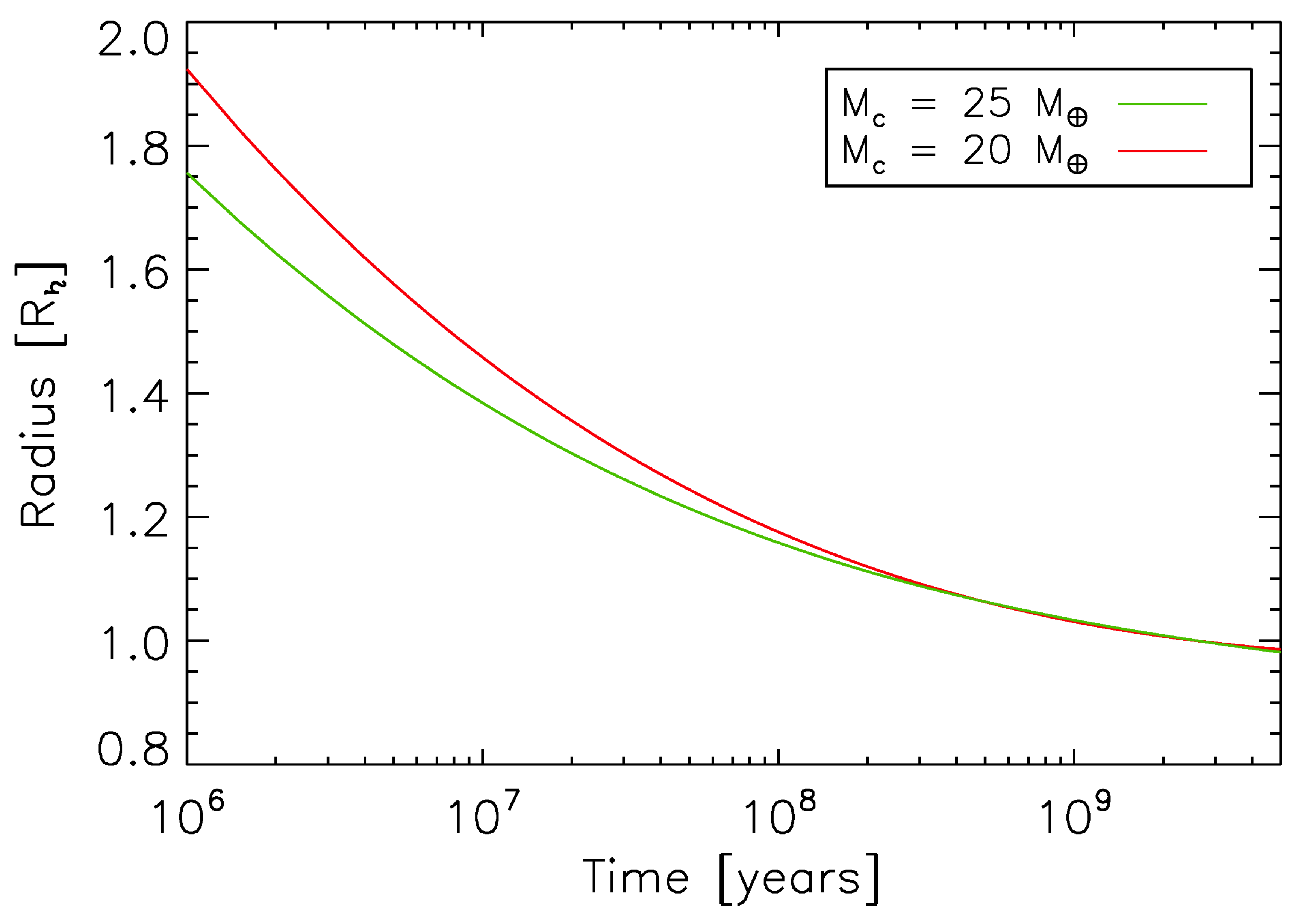}
\caption{Radius of a Saturn-equivalent planet, as a function of time since the planet's formation, for an assumed core of $25 M_\oplus$ (green line) and $20 M_\oplus$ (red line). Data for the $25 M_\oplus$ has been extracted from Figure 5B of \cite{fortney07}. Data for the $20 M_\oplus$ was provided by W. Fortney for \cite{canup10}.}
\label{figSaturnContraction}
\end{center}
\end{figure}

We consider the evolution of a ring formed soon after the end of Saturn's gas accretion, at which time the planet will still be substantially larger than its current size due to the energy of its formation.  To estimate the physical radius of Saturn, we use results from \cite{fortney07}. The green dotted-dashed line in their Figure 5B represents the evolution of a 0.3 Jupiter mass planet, which is about the mass of Saturn, assuming a $25M_{\oplus}$ core. A fit to this data is shown as the green line in Figure \ref{figSaturnContraction}. A core mass of $\sim 20 M_\oplus$ \citep{hubbard09} results in the red curve (data provided by W. Fortney for \cite{canup10}), an approximate fit to which is $R(t) = A_0 + A_1log(t) + A_2log(t)^2 + A_3log(t)^3$, where $R$ is in units of $R_{\saturn}$, $t$ is in years, $A_0\approx9.576$, $A_1\approx -2.418$, $A_2\approx 0.231$, and $A_3\approx -0.0075$.  

\begin{figure}[!h]
\begin{center}
\includegraphics[width=8.5cm]{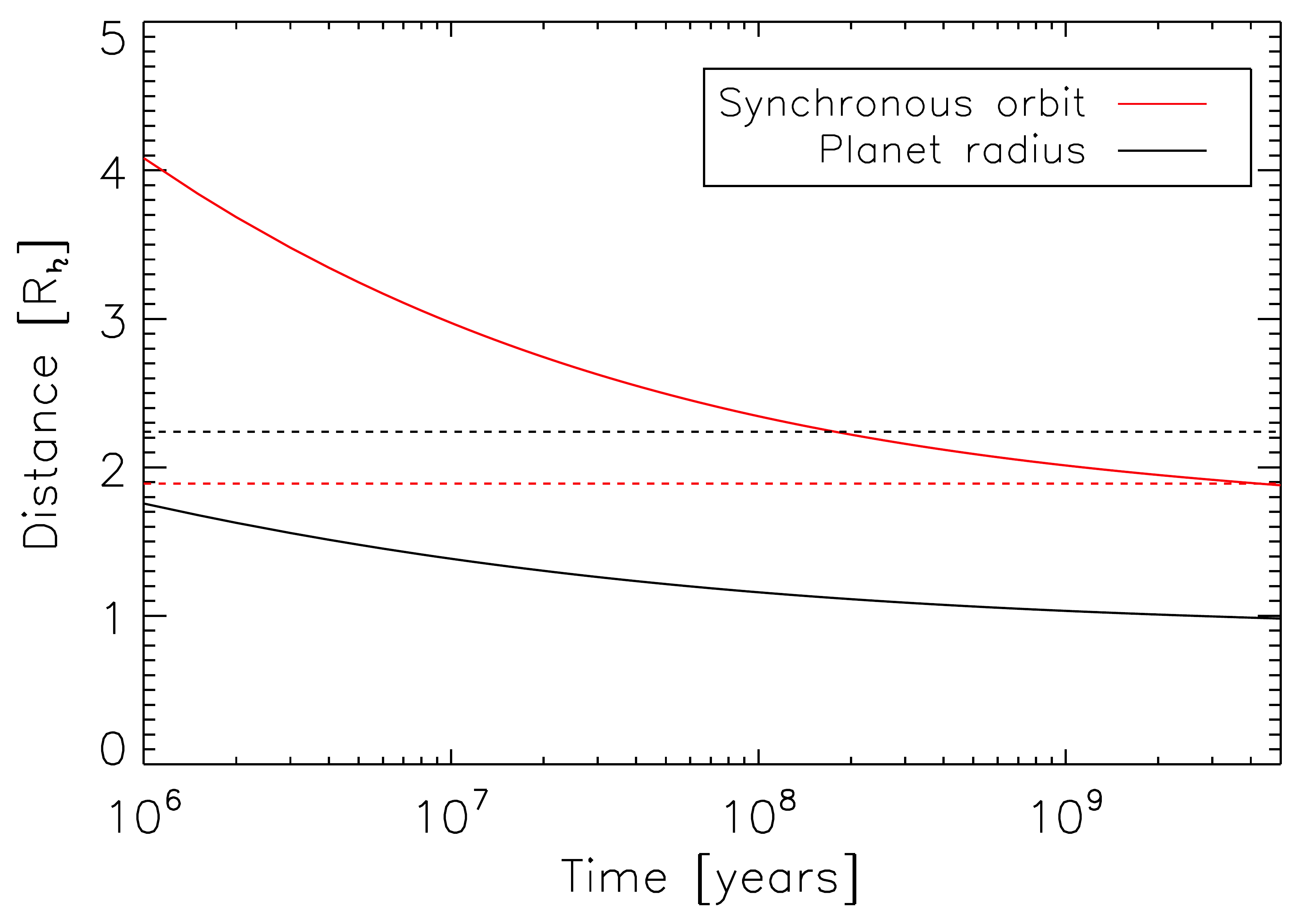}
\caption{Position of the synchronous orbit (red line) and physical radius of the planet (black line) as a function of time since Saturn's formation. The horizontal black dashed line is the position of the Roche limit. While today the synchronous orbit lies at $\approx \SI{1.89}{R_{\saturn}}$ (red dashed line), it was exterior to the Roche limit for $\sim 10^8$ yr, and remains exterior to its current position for $\sim 10^9$ yr.}
\label{figSynchronousOrbit}
\end{center}
\end{figure}

From conservation of its spin angular momentum, one can estimate Saturn's early spin rate as a function of its physical radius and moment of inertia.  We assume an early moment of inertia constant comparable to that of current Saturn, $K_{\saturn} \approx 0.23$ \citep{helled11,nettelmann13}, see Appendix.  The resulting predicted evolution of the synchronous orbit with time is shown in Figure \ref{figSynchronousOrbit}.  While currently $a_{sync}$ lies well inside the Roche limit, for the first $\sim 10^9$ yr of Saturn's history synchronous orbit is shifted outward due to the slower rotation of the planet. For moons near the Roche limit, there will thus be a competition between the negative torque due to tides (causing orbital contraction) and the positive torque due to resonant interactions with the rings (causing orbial expansion).  For $Q \ge 10^4$ and $\sigma \ge 10^3$ g cm$^{-3}$, the latter are much stronger, allowing spawned satellites to evolve away from the rings.

\subsection{Simulation parameters}

Table \ref{table_simu_parameters} lists parameters for our 12 baseline cases.   We consider $R_p = $ 1.3, 1.4 or $\SI{1.5}{R_{\saturn}}$, with corresponding planetary rotational periods of 17.9, 20.7 and $\SI{23.8}{hours}$, respectively.  For orbits far beyond synchronous, $|\omega| >> n$ and Saturn's tidal time lag is approximately $\Delta t \approx 1/(2|\omega|Q)$.  We set $\Delta t$ using this expression so that $Q = 10^4$. We consider initial ring masses between $\SI{3e21}{\kilogram}$ and $\SI{1.1e22}{\kilogram}$, motivated by models of tidal stripping from a Titan-sized satellite \citep{canup10}.  We consider ice ring particles with density $\rho_r = \SI{0.9}{g.cm^{-3}}$, which sets the Roche limit at $a_R \approx \SI{2.24}{R_{\saturn}}$.  Our initial ring extends from the planet's physical radius $R_p$ to The Roche limit. It is possible that the ring may have been more concentrated initially, but it would rapidly viscously spread \citep{salmon10}. 

We complete four sets of 12 baseline simulations \ref{table_simu_parameters}.  The first and second sets assume no pre-existing exterior satellites, and consider either no satellite tides ($\mathcal{A}=0$; ``Set A'') or strong satellite tides ($\mathcal{A}= 10^3$; ``Set B''). The third and fourth sets include Dione and Rhea at their current locations, intended to represent the earlier formation of these outer moons as direct accretional products from the Saturnian subnebula, both with no satellite tides (``Set C'') and with strong satellite tides (``Set D'').

\begin{deluxetable}{l c c c c c c}
\tabletypesize{\footnotesize}
\tablewidth{0pt}
\tablecolumns{7}
\tablecaption{Simulation parameters. \label{table_simu_parameters}}
\tablehead{ \colhead{Run} & \colhead{$R_p$} & \colhead{$T$} & \colhead{$a_{sync}$} & \colhead{$a_R$} & \colhead{$\Delta t$} & \colhead{$M_d$}\\
  & \colhead{$\left(R_{\saturn}\right)$} & \colhead{(hours)} & \colhead{$\left(R_p\right)$} & \colhead{$\left(R_p\right)$} & \colhead{(seconds)} & \colhead{$\left(\SI{e-5}{M_{\saturn}}\right)$}}
\startdata
1  & 1.5 & 23.8 & 2.19 & 1.50 & 0.68 & 0.5 \\
2  & 1.5 & 23.8 & 2.19 & 1.50 & 0.68 & 1 \\
3  & 1.5 & 23.8 & 2.19 & 1.50 & 0.68 & 1.5 \\
4  & 1.5 & 23.8 & 2.19 & 1.50 & 0.68 & 2 \\
5  & 1.4 & 20.7 & 2.14 & 1.60 & 0.59 & 0.5 \\
6  & 1.4 & 20.7 & 2.14 & 1.60 & 0.59 & 1 \\
7  & 1.4 & 20.7 & 2.14 & 1.60 & 0.59 & 1.5 \\
8  & 1.4 & 20.7 & 2.14 & 1.60 & 0.59 & 2 \\
9  & 1.3 & 17.9 & 2.09 & 1.73 & 0.51 & 0.5 \\
10 & 1.3 & 17.9 & 2.09 & 1.73 & 0.51 & 1 \\
11 & 1.3 & 17.9 & 2.09 & 1.73 & 0.51 & 1.5 \\
12 & 1.3 & 17.9 & 2.09 & 1.73 & 0.51 & 2 \\
\enddata
\tablecomments{$R_p$ is the planet's mean physical radius in units of Saturn current mean radius $R_{\saturn}=\SI{58232}{km}$. $T$ is the spin period of the planet. $a_{sync}$ and $a_R$ are the position of the synchronous orbit and of the Roche limit, respectively, in units of the planet's physical radius. $\Delta t$ is the tidal lag. $M_d$ is the disk's initial mass.}
\end{deluxetable}

\section{Results}
\subsection{General accretion dynamics}
\begin{figure}
\begin{center}
\includegraphics[width=8cm]{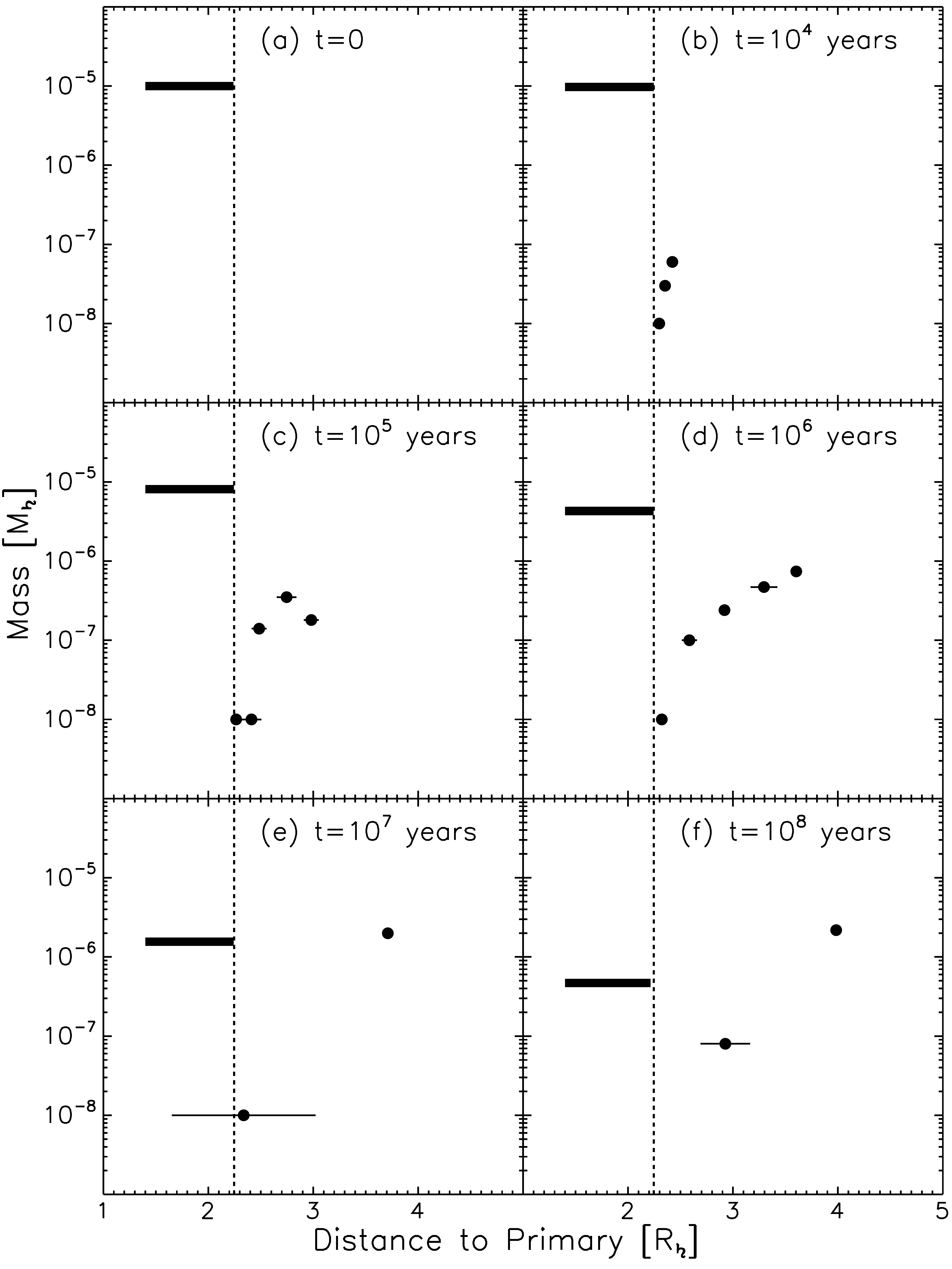}
\caption{Snapshot of the system in Run 6a at different times of evolution. The vertical dashed line at $\approx 2.24R_{\saturn}$ is the Roche limit. The thick black horizontal line is the Roche-interior ring, whose inner edge is at the planet's surface at $R_p = 1.4 R_{\saturn}$. The black dots represent the satellites formed from the disk, with the thin horizontal lines representing their pericenter and apocenter.}
\label{figSimuPannels}
\end{center}
\end{figure}

Figure \ref{figSimuPannels} shows the system at different evolution times for the first $\SI{e8}{years}$, for Run 6A. As the ring spreads, it starts producing new moonlets that, through resonant interaction, confine the ring inside the Roche limit. In turn, they recoil and the ring is progressively freed to viscously spread again. When a moonlet reaches $\ge \SI{3.56}{R_{\saturn}}$, its 2:1 Lindblad resonance lies beyond $a_R$ and it thus stops directly interacting with the ring.

\begin{figure}[!h]
	\begin{center}
		\includegraphics[width=8cm]{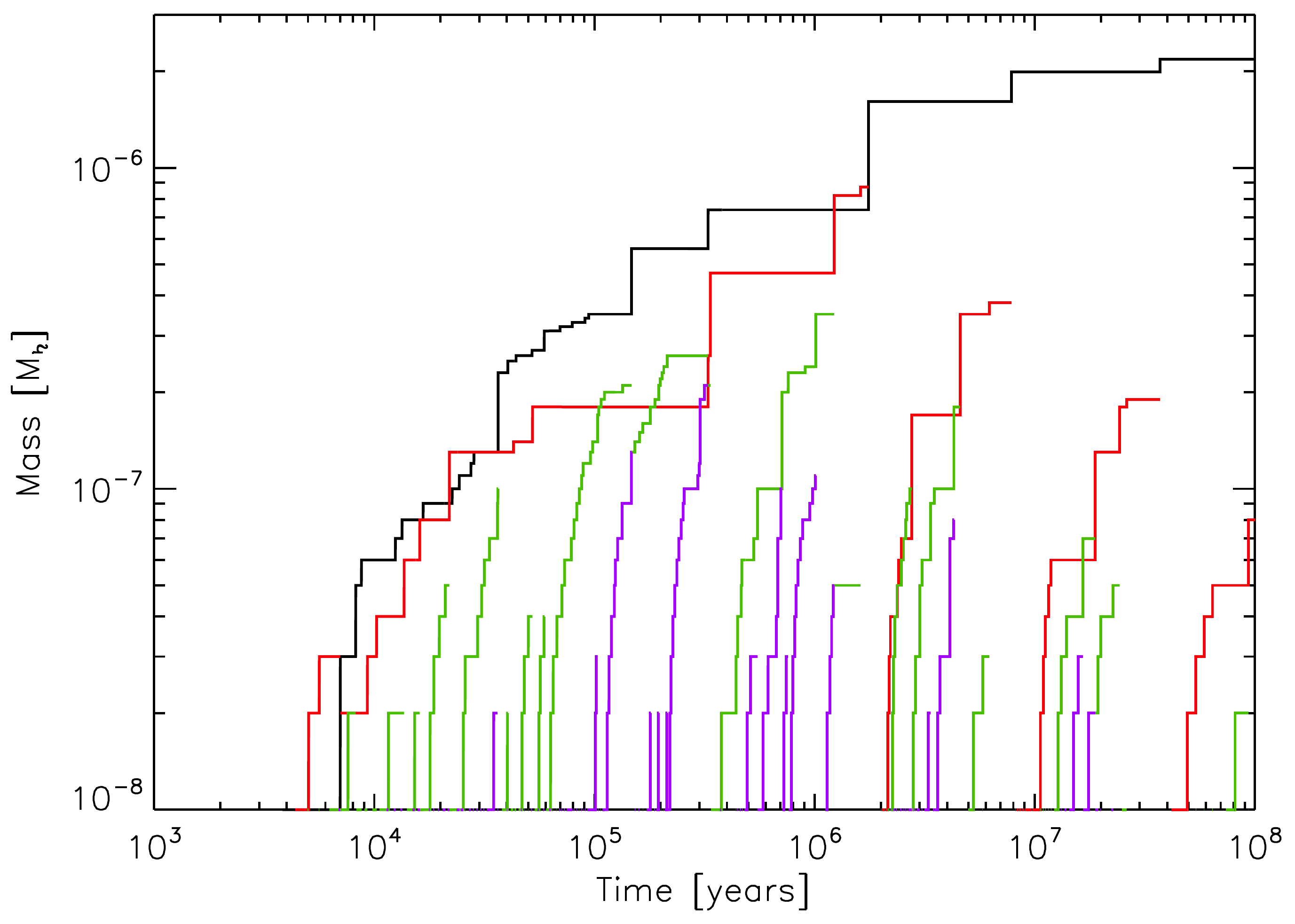}
		\caption{Evolution of the masses of multiples bodies in Run 6A. For readability, only the 4 oldest bodies at any given time are shown. Colors correspond to an index in our output mass array (see text for details). Satellites grow initially by direct accretion of moonlets spawned at the Roche limit (similar to the ``discrete regime'' of \cite{crida12}), and later by merger of grown satellites (the ``pyramidal regime'' of \cite{crida12}). After about $10^4$yr, there are between 1 and 7 mid-sized moons at any given time.}
		\label{figSatMasses}
	\end{center}
\end{figure}

When a moonlet is spawned at the Roche limit in the presence of exterior satellites, it will encounter their mean motion resonances (MMRs) as it recoils outward due to ring torques. Initially, the ring is massive enough that its torques typically cause moonlets to recoil too rapidly for capture into resonance. As a result, as inner moonlets expand outward they can have close encounters with outer satellites that can result in a merger and the growth of increasing massive moons. Figure \ref{figSatMasses} shows the masses of various objects in Run 6A, as a function of time. Colors correspond to indexes in our output mass array: black for moonlet \#1 (which formed first and is the oldest), red for moonlet \#2, green for moonlet \#3, and purple for moonlet \#4. Other bodies may be present at times but have not been plotted for readability. Color changes occur when two objects merge. For example, at $t\approx \SI{e6}{years}$, moonlet \#3 (in green) merges with \#2 (in red), and then moonlet \#4 becomes the new moonlet \#3, changing color from purple to green.

Our simulations display the general behavior predicted in \cite{crida12}. Initially a moon spawned near the Roche limit directly accretes small ring material as it spreads across the Roche limit; this is defined as the ``continous regime'' of growth \citep{crida12}.  As the moon rapidly recoils outward due to ring torques, its separation from the ring edge becomes large enough for a second inner satellite can begin to grow near the Roche limit. This second satellite also recoils outward and is eventually accreted by the outer satellite.  This process repeats so long as the first satellite is relatively close to the ring's edge, with the first satellite growing at the same average rate as in the continous regime, only through larger discrete steps; accordingly this mode of growth is called the ``discrete regime'' \citep{crida12}. Finally as the first satellite continues to evolve outward it can become distant enough that it can no longer directly accrete a moon spawned from the ring, and a system of three or more moons results. Mergers are then characterized by collisions between similar-mass bodies in the so-called ``pyramidal regime'' \citep{crida12}. The transition between the discrete and pyramidal regimes is predicted to occur when a moonlet of mass $m$ reaches a distance $r$ such that $r - 2r_H > r_c$, where $r_H= r\left(m/(3M_{\saturn})\right)^{1/3}$ is the moonlet's Hill radius and $r_c=a_R\left(8.4M_{disk}/M_{\saturn}+1\right)$ \citep{crida12}. In our simulations, $M_{disk}/M_{\saturn}\sim 10^{-5}$ such that moonlets should transition to the pyramidal regime after only little outward migration. This is indeed observed but at somewhat larger distances than predicted by the above expression. We find that several moonlets can be in the discrete regime at the same time, e.g. black and red curves before $10^5$ years on Figure \ref{figSatMasses}. Also, we find that a younger moonlet can grow larger than an older one, e.g. black and red curves around $10^5$ and $10^6$ years in Figure \ref{figSatMasses}. Configurations with an inner moon that is larger than an outer one are however generally only transient, as merging events between large objects eventually produce a system with larger satellites at larger distances, consistent with the \cite{crida12} expectations.

As the ring mass decreases due to mass loss on the planet and by formation of moonlets at the Roche limit, two things can be noted. First, the ring's viscosity decreases and the time needed for the ring to spread back out to the Roche limit increases, such that the time between the spawning of new moonlets lengthens. Second, the Lindblad resonant torque becomes weaker and the orbital expansion of inner bodies slows, such that they can be captured into MMRs with outer bodies. When this happens, the inner object continues to recoil outward as it is torqued by the ring, and in turn it drives the outer object outward as well due to the resonant configuration. This process allows for a transfer of angular momentum from the ring to outer objects that do not themselves have direct resonant interactions with the ring. This is a key process not included in the \cite{charnoz11} and \cite{crida12} models.

\begin{figure}[!h]
	\begin{center}
		\includegraphics[width=8.5cm]{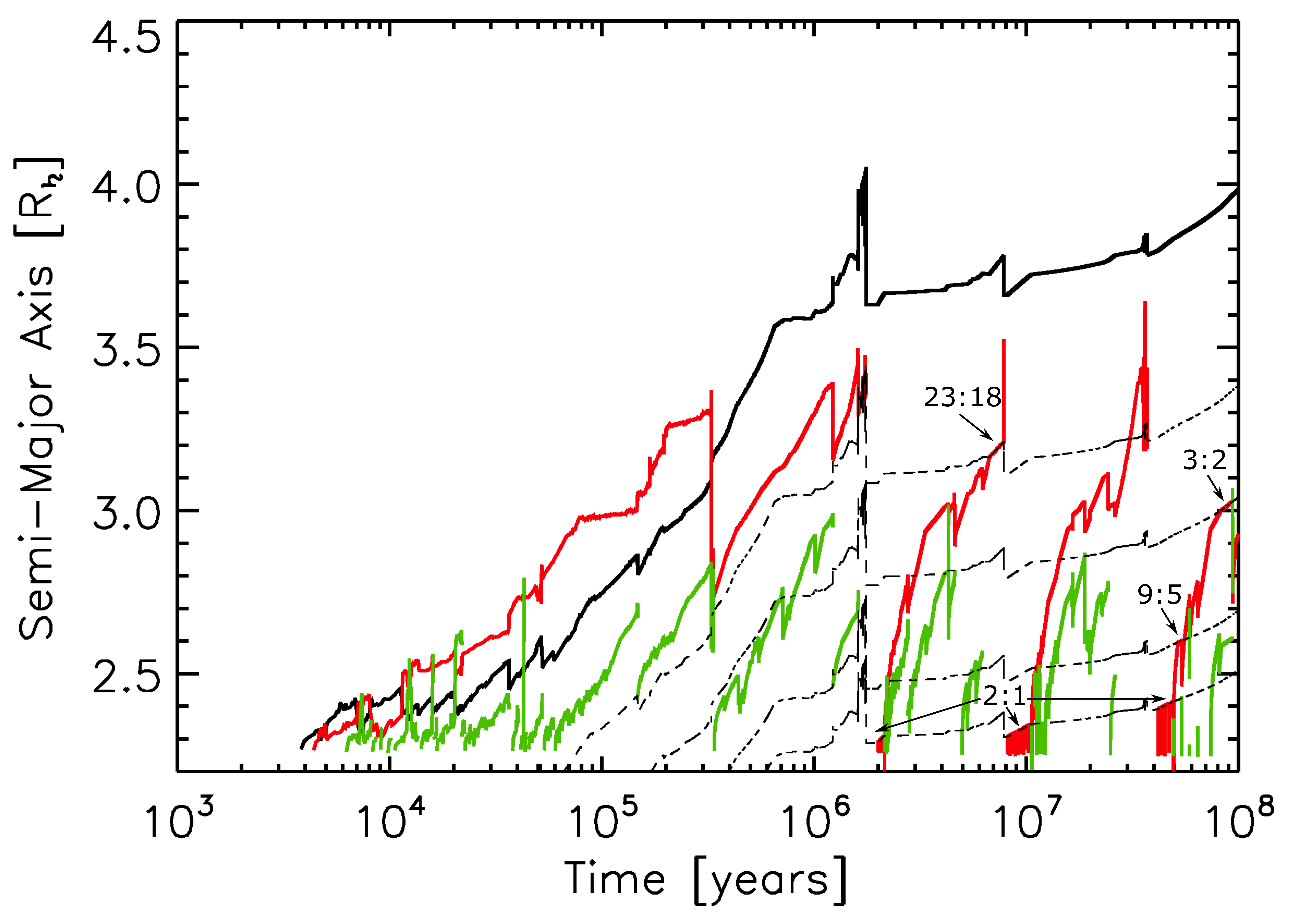}
		\caption{Evolution of the semi-major axis of a satellite (black curve), and position of some of its MMR (black dashed lines) in Run 6A. The red and green curves represent the semi-major axis of secondary bodies spawned at the Roche limit. Moonlets stop interacting with the disk when they reach $\sim3.5R_{\saturn}$ as their 2:1 Lindblad resonance lies outside the Roche limit. However, by capturing into MMR inner objects which are themselves still interacting with the disk, outer objects can reach larger distances on timescales short compared to what could be achieved by tidal interactions with the planet.}
		\label{figSMAEvolution}
	\end{center}
\end{figure}

The black line in Figure \ref{figSMAEvolution} shows the evolution of an object's semi-major axis in Run 6A. The object is initially spawned at the Roche limit at $t \sim \SI{e4}{years}$ and moves outward due to resonant interactions with the rings. This object is not massive enough to confine the rings so that secondary objects are subsequently spawned and also recoil (red and green curves in Figure \ref{figSMAEvolution}). As they catch up with the outer object, mergers can occur, and cause the outer object's semi-major axis to decrease somewhat due to the accreted object having a lower specific angular momentum \citep{salmon12}. At $10^8$ years, the two most massive satellites in Run 6A have masses of $\SI{2.18e-6}{M_{\saturn}}$ and $\SI{8e-8}{M_{\saturn}}$, and semi-major axes of $\SI{3.94}{R_{\saturn}}$ and $\SI{2.78}{R_{\saturn}}$, respectively. They are similar to Tethys and Enceladus in mass, but their semi-major axes are smaller. Further expansion will be achieved over longer timescales due to tides and/or MMR interactions. This run did not produce a Mimas-equivalent satellite within $\SI{e8}{years}$. 

Close encounters between satellites do not always result in a merger as assumed  in \cite{crida12}, but can instead lead to scattering, inward or outward. Such an event can be seen at $\SI{e4}{years}$ in Run 6A (Figures \ref{figSatMasses} and \ref{figSMAEvolution}, black and red lines). This leads to an orbital architecture in which the outermost body is less massive than the one immediately inside. This situation is transient as the two moons re-exchange orbits, at $\sim \SI{3e5}{years}$ in this case.

\begin{figure}[!h]
	\begin{center}
		\includegraphics[width=8.5cm]{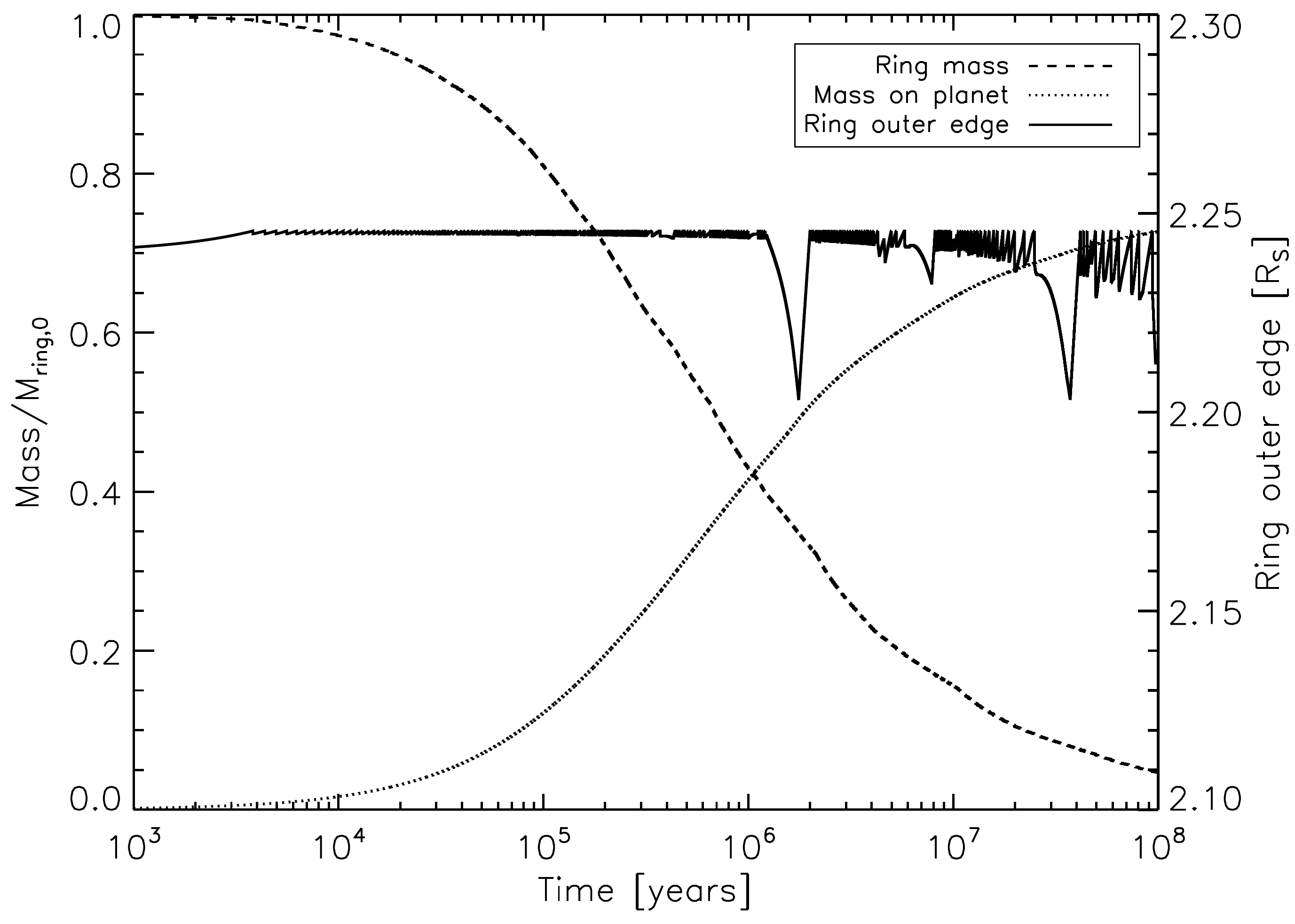}
		\caption{Ring's outer edge (solid line), mass (dashed line) and mass fallen onto the planet (dotted line), for Run 6A. Masses are normalized to the initial mass of the ring $M_{ring,0}$. Due to the constant confinement of the ring by growing satellites, about $70\%$ of the ring's material is lost onto the planet. At times, a satellite's torque can surpass the ring's viscous torque, resulting in a prolonged contraction of the ring's outer edge.}
		\label{figDiskInfo}
	\end{center}
\end{figure}

Figure \ref{figDiskInfo} shows the evolution of the rings in Run 6A: position of the outer edge (solid line), mass (dashed line), and mass fallen on the planet (dotted line). When a satellite is spawned at the ring's outer edge, resonant interactions cause the latter to slightly contract inside the Roche limit. Early on, the ring is still massive enough that its viscous torque is greater than the resonant torque from outer satellites, and the ring viscously spreads outward. As satellites grow larger through mutual collisions, and as the ring's mass decreases, the resonant torque can at times overwhelm the viscous torque, resulting in a prolonged contraction of the ring's outer edge (Figure \ref{figDiskInfo}, solid line at e.g. $\SI{e6}{years}$). As the confining satellite's orbit expands due to the resonant torque (Figure \ref{figSMAEvolution}), the torque decreases both because the distance between the satellite and disk edge increases and because resonances with the satellite move outward with it and migrate out of the ring. Eventually, the ring viscously spreads outward again.

Whenever the ring is confined inside the Roche limit, it continues to lose mass onto the planet through its inner edge, while not providing any additional mass to outer satellites. As a result, in this simulation more than $70 \%$ of the ring's mass is lost onto the planet. Mimas, Enceladus and Tethys's masses total $\SI{1.35e-6}{M_{\saturn}}$, so that a ring initially at least 3 times as massive with $ > \SI{4e-6}{M_{\saturn}}$ would be necessary to produce these objects. In this run the rings still contain $\sim \SI{2.5e-7}{M_{\saturn}}$ at $10^8$ years, which is about 4 times the mass of Mimas. Formation of a Mimas-equivalent may thus occur on longer timescales (see Section \ref{Section_Simulations1d9}).

\subsection{Set A: No satellite tides}
\begin{figure*}[!h]
\begin{center}
\includegraphics[width=18cm]{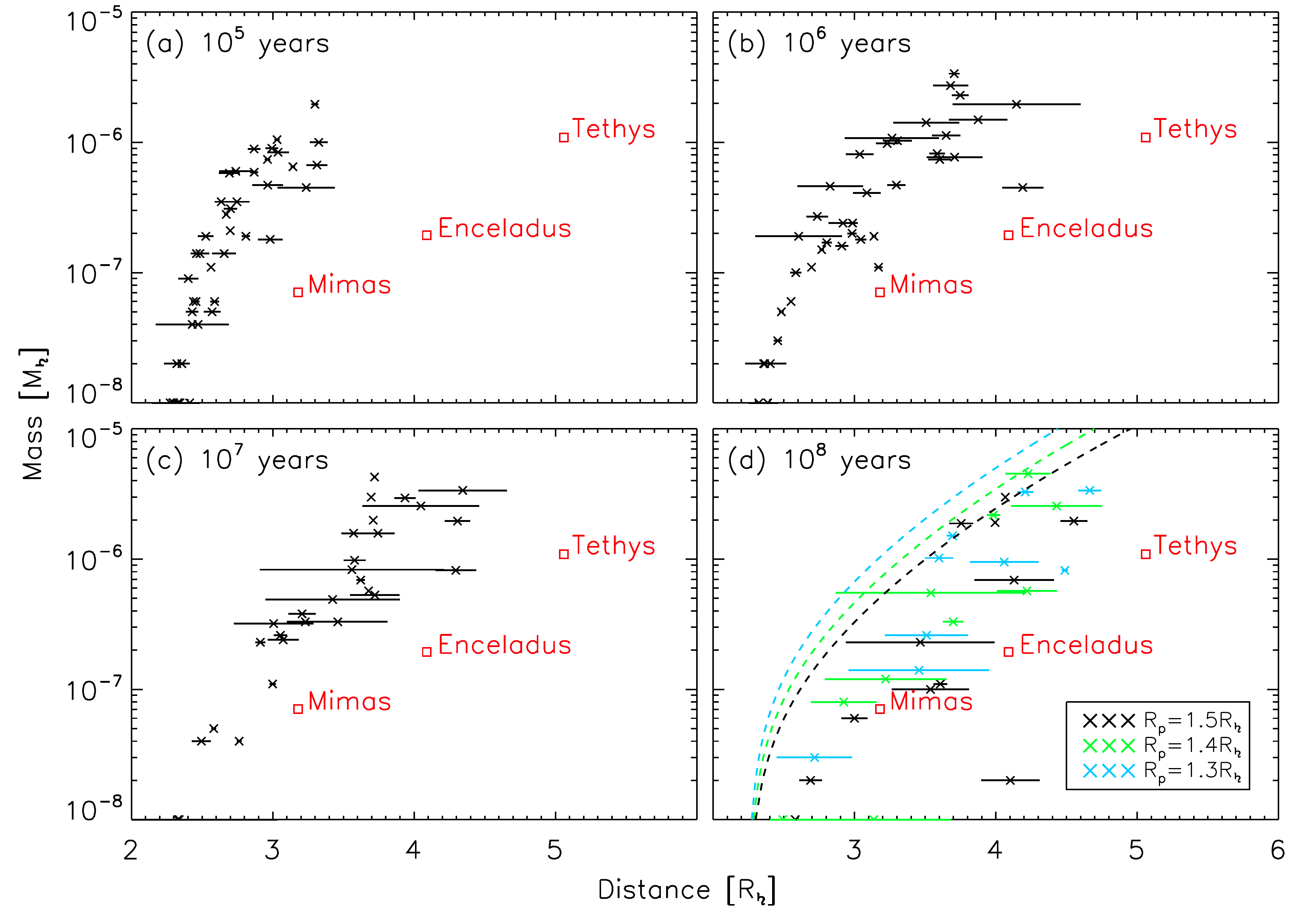}
\caption{Distribution of satellites in our Set A simulations at evolution times of $10^5$, $10^6$, $10^7$ and $\SI{e8}{years}$. The red squares represent Mimas, Enceladus and Tethys. Horizontal lines show pericenter and apocenter for each satellite. In panel (d), colors separate satellites based on the assumed radius of the planet. Dashed lines represent the distance that a satellite of a given mass, originating at the Roche limit, would have reached solely due to tides per Eq \ref{equ_smalim}. Nearly all satellites lie to the right of these lines, because they have also orbitally expanded due to disk torques and MMRs.}
\label{figSatDistr_noDR}
\end{center}
\end{figure*}

Figure \ref{figSatDistr_noDR} shows the distribution of satellites obtained in our $10^8$-year simulations without satellites tides and without pre-existing Dione and Rhea at different times of evolution. Spawned satellites have masses broadly comparable to those of Mimas, Enceladus and Tethys. After $\sim \SI{e8}{years}$, some objects have nearly reached the position of Tethys' orbit, primarily due to ring torques and capture into MMRs. The evolution of the semi-major axis of a satellite of mass $m$ due to tides can be estimated by \citep{burns77}:
\begin{equation}
\frac{da}{dt}=\frac{3k_2}{Q}\frac{m}{M_{\saturn}}\sqrt{\frac{GM_{\saturn}}{R_p}}\left(\frac{R_p}{a}\right)^{11/2},
\end{equation}
which can be integrated to give 
\begin{equation}
a(t)=R_p\left[\frac{13}{2}\frac{3k_2}{Q}\frac{m}{M_{\saturn}}\sqrt{\frac{GM_{\saturn}}{R_p^3}}t+\left(\frac{a_0}{R_p}\right)^{13/2}\right]^{2/13},
\label{equ_smalim}
\end{equation}
where $a_0$ is the satellite's initial semi-major axis. Panel (d) in Figure \ref{figSatDistr_noDR} shows $a(m)$ from Eq. \ref{equ_smalim} for the three values assumed for $R_p$, with $Q=10^4$ and $t=10^8$ years; the satellites produced in our simulations are shown with the same color scheme. Most satellites are to the right of their corresponding curve, indicating that they have orbits larger than expected due solely to tides. Satellites beyond $\sim\SI{3.55}{R_{\saturn}}$ have no resonances in the rings, and their orbital expansion beyond the dashed curves has been achieved by trapping of inner satellites into MMRs, i.e., by indirect angular momentum transport from the ring.

Table \ref{table_seta_results} shows results from the Set A simulations.  On average, they yield $2.6 \pm 0.7$ final satellites at $\SI{e8}{years}$; at that time, the rings have an average mass of $\SI{3.83e-7}{M_{\saturn}}$, $\SI{4.78e-7}{M_{\saturn}}$ and $\SI{5.96e-7}{M_{\saturn}}$, for the runs with a planet radius of 1.5, 1.4 and $\SI{1.3}{R_{\saturn}}$ respectively. Runs with larger planets have a lower ring mass at a given time because the flux onto the planet has been larger for a given initial ring mass. For each value of $R_p$, the ring mass is very similar at $t=10^8$ even though the initial ring masses vary by a factor of 4. We find that about $20\%$ of the ring's initial mass is incorporated into satellites, with a slightly higher fraction for smaller values of $R_p$. The average angular momentum of our satellites is higher than that of current Mimas, Enceladus and Tethys (though standard deviation is important), mostly because our Enceladus and Tethys analogs tend to be a factor of a few times more massive than the current moons. 

In all cases, sufficient mass and angular momentum remains in the rings to spawn additional satellites over longer timescales. The average angular momentum left in the rings is $\SI{4.4e32}{kg.m^2.s^{-1}}$, $\SI{5.49e32}{kg.m^2.s^{-1}}$ and $\SI{6.74e32}{kg.m^2.s^{-1}}$, for the runs with a planet radius of 1.5, 1.4 and $\SI{1.3}{R_{\saturn}}$ respectively.  This is a few times larger than the angular momentum necessary to bring a Tethys-mass satellite from 4 to $\SI{5}{R_{\saturn}}$ ($\SI{2.2e32}{kg.m^2.s^{-1}}$). It appears then likely that the distribution of satellites will continue to evolve significantly on longer timescales, a point we return to in Section 4.

Eccentricities and inclinations of objects trapped in MMRs increase as the inner object is driven outward by the disk. The run shown in Figure \ref{figSimuPannels} shows satellites with small eccentricities but is a bit of an outlier in this regard compared to the satellites formed in the whole Set. Damping of eccentricities occurs when objects collide, but significant values are generally reached in the absence of satellite tides. Bigger objects have on average smaller eccentricities and inclinations, since these objects have experienced more collisions, and it is more difficult to excite $e$ and $i$ for a bigger satellite. At $t=\SI{e8}{years}$, satellites with masses smaller than $\SI{e-6}{M_{\saturn}}$ have $\langle e \rangle \sim \num{0.073} \pm 0.05$ (median is $\num{0.068}$), while those with masses larger than $\SI{e-6}{M_{\saturn}}$ have $\langle e \rangle \sim \num{0.022} \pm 0.02$ (median is $\num{0.018}$). Inclinations of satellites also become substantial, with an average of $\langle i \rangle \sim \SI{3.32}{\degree} \pm \SI{3.62}{\degree}$ (median is $\SI{1.95}{\degree})$ for small satellites, and $\langle i \rangle \sim \SI{2.71}{\degree} \pm \SI{5.14}{\degree}$ (median is $\SI{0.23}{\degree})$ for large satellites.

\begin{deluxetable}{l c c c}
	\tabletypesize{\footnotesize}
	\tablewidth{0pt}
	\tablecolumns{4}
	\tablecaption{Set A data at $t=10^8$ years. \label{table_seta_results}}
	\tablehead{ & \colhead{$R_p=1.5R_{\saturn}$} & \colhead{$R_p=1.4R_{\saturn}$} & \colhead{$R_p=1.3R_{\saturn}$}}
	\startdata
	$<M_{ring}/M_{MET}>$  & $0.28 \pm 0.01$ & $0.36 \pm 0.01$ & $0.44 \pm 0.001$\\
	$<L_{ring}/L_{MET}.$  & $0.18 \pm 0.01$ & $0.22 \pm 0.01$ & $0.27 \pm 0.002$\\
	$<M_{sats}/M_{MET}>$ & $1.85 \pm 0.96$ & $2.03 \pm 1.1$ & $2.11 \pm 1.1$\\
	$<M_{sats}/M_{ring,0}>$ & $0.2 \pm 0.01$ & $0.22 \pm 0.01$ & $0.23 \pm 0.01$\\
	$<L_{sats}/L_{MET}>$ & $1.7 \pm 0.89$ & $1.88 \pm 1.04$ & $1.98 \pm 1.02$\\
	\enddata
	\tablecomments{Average values for our Set A runs at $t=10^8$ years. $M_{ring}$ is the rings' mass, $M_{MET}$ is the total mass of Mimas, Enceladus and Tethys, $L_{ring}$ is the rings' angular momentum, $M_{sats}$ is the total mass of the satellites in a given Run, $M_{ring,0}$ is the rings' initial mass, $L_{sats}$ is the angular momentum of the satellites in a given Run, and $L_{MET}$ is the total angular momentum of Mimas, Enceladus and Tethys.}
\end{deluxetable}

\subsection{Set B: Strong satellite tides}
\begin{figure}
	\begin{center}
		\includegraphics[width=8cm]{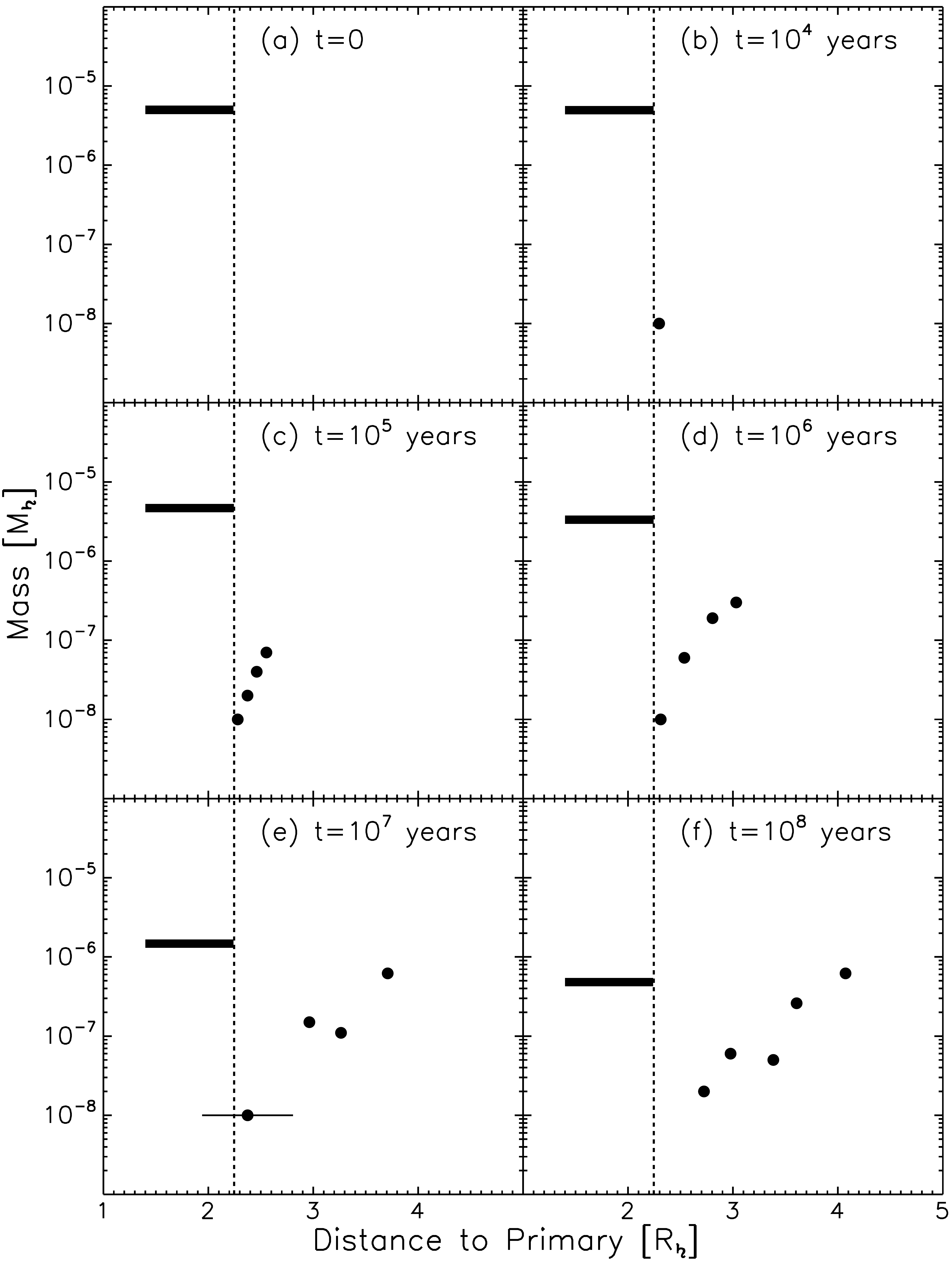}
		\caption{Snapshot of the system in Run 5B at different times of evolution. The vertical dashed line at $\approx 2.24R_{\saturn}$ is the Roche limit. The thick black horizontal line is the Roche-interior ring, whose inner edge is at the planet's surface at $R_p = 1.4 R_{\saturn}$. The black dots represent the satellites formed from the disk, with the thin horizontal lines representing their pericenter and apocenter.}
		\label{figSimuPannelsb}
	\end{center}
\end{figure}

We perform a second set of simulations with the same initial parameters as in Table \ref{table_simu_parameters} but including satellite tides with $\mathcal{A}=1000$. Figure \ref{figSimuPannelsb} shows a system at different times of evolution. This case's evolution is similar to that in Figure \ref{figSimuPannels}, the main difference being the smaller eccentricities at all times, a direct consequence of satellite tides. Compared to the case without satellite tides, the largest satellite is smaller. This is due to a factor of 2 difference in the mass of the disk, which results in less massive satellites being formed and weaker orbital expansion rates, which overall decreases the delivery rate of material beyond the Roche limit.

\begin{figure*}[!h]
	\begin{center}
		\includegraphics[width=18cm]{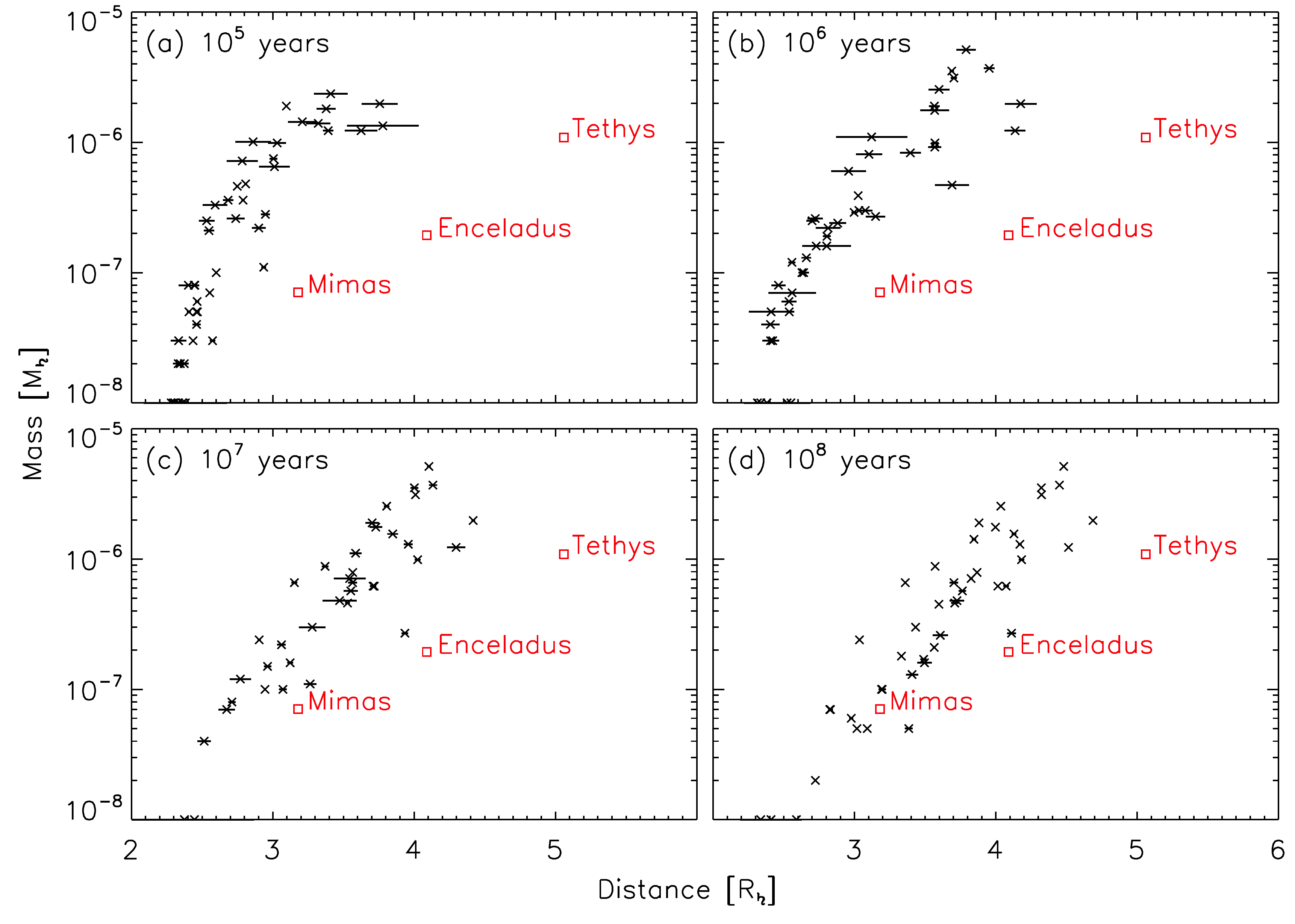}
		\caption{Distribution of satellites in our Set B simulations including satellites tides with $\mathcal{A}=1000$ at evolution times of $10^5$, $10^6$, $10^7$ and $\SI{e8}{years}$. The red squares represent the current Mimas, Enceladus and Tethys. Horizontal lines show pericenter and apocenter of the satellites. Compared to Set A, here satellite tides efficiently damp eccentricities.}
		\label{figSatDistr_noDR_SatTides}
	\end{center}
\end{figure*}

The overall results at $t=\SI{e8}{years}$ for the Set B runs are given in Table \ref{table_setb_results}. Figure \ref{figSatDistr_noDR_SatTides} shows the distribution of satellites at different times of evolution. The lower eccentricities of the satellites noted in Run 5B (Figure \ref{figSimuPannelsb}) can be observed across all runs. While without satellites tides some moons in Set A also have low eccentricities at $t=\SI{e8}{years}$, they went through a phase of high values before they were damped by accretional collisions (Figure \ref{figSatDistr_noDR}).

\begin{figure}[!h]
	\begin{center}
		\includegraphics[width=8.5cm]{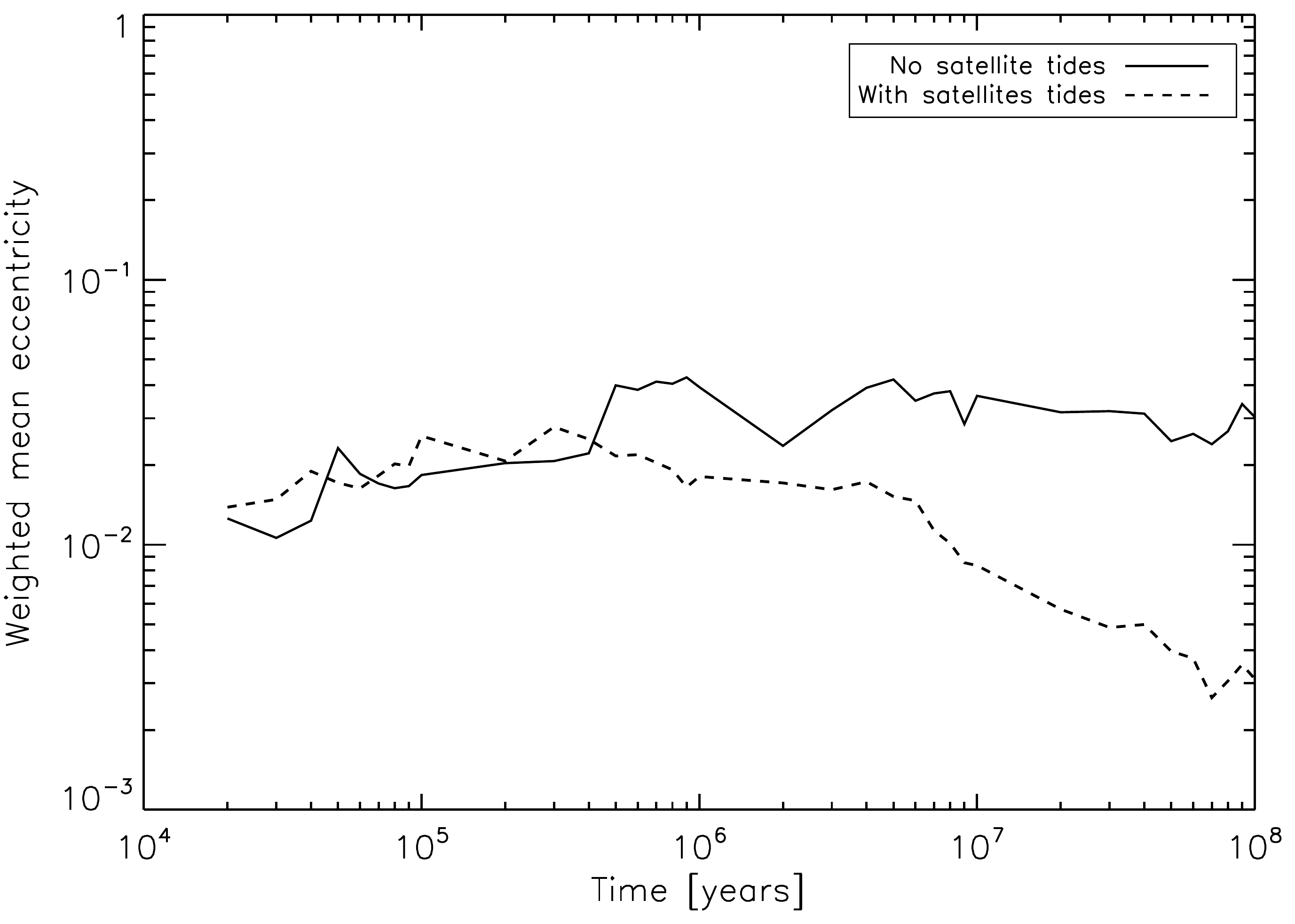}
		\caption{Mean eccentricity of formed satellites weighted by the satellite's mass, as a function of time, in the case of no satellites tides (solid line; Set A) and including satellites tides (dashed line; Set B).}
		\label{fig_ecc_comp}
	\end{center}
\end{figure}

Figure \ref{fig_ecc_comp} shows the mean eccentricities of the satellites as a function of time, weighted by the mass of the satellites, for the Set A vs. Set B runs. The initial eccentricities of a fragment spawned at the Roche limit in our simulations is set to be approximately the ratio of the fragment's escape velocity to the local orbital velocity \citep{lissauer93}, which is $\sim 10^{-3}$ for a $\SI{e-8}{M_{\saturn}}$ mass fragment.   Set B runs with tidal dissipation in the satellites (dashed line) experience efficient damping occurs on a $\ge 10^6$ year timescale, resulting in much smaller average eccentricities at \SI{e8}{years}, compared to our runs without satellites tides (solid line).
 
Compared with Set A, there is a somewhat higher average fraction of the ring's initial mass incorporated into satellites, which in Set B approaches 30\%. On average, the Set B simulations have $3.75 \pm 1.05$ satellites at $\SI{e8}{years}$. A larger final number of spawned moons is a consequence of the satellites' lower eccentricities: their pericenter is larger, and they do not ``sweep'' the region close to the rings as efficiently, allowing more small moons to form and survive. The combination of more mass put into satellites and lower eccentricities contributes to a higher total angular momentum in spawned moons compared with Set A.

\begin{deluxetable}{l c c c}
	\tabletypesize{\footnotesize}
	\tablewidth{0pt}
	\tablecolumns{4}
	\tablecaption{Set B data at $t=10^8$ years. \label{table_setb_results}}
	\tablehead{ & \colhead{$R_p=1.5R_{\saturn}$} & \colhead{$R_p=1.4R_{\saturn}$} & \colhead{$R_p=1.3R_{\saturn}$}}
	\startdata
	$<M_{ring}/M_{MET}>$  & $0.22 \pm 0.07$ & $0.33 \pm 0.04$ & $0.45 \pm 0.005$\\
	$<L_{ring}/L_{MET}>$  & $0.13 \pm 0.04$ & $0.20 \pm 0.02$ & $0.27 \pm 0.004$\\
	$<M_{sats}/M_{MET}>$ & $2.19 \pm 1.48$ & $2.47 \pm 1.69$ & $2.69 \pm 1.86$\\
	$<M_{sats}/M_{ring,0}>$ & $0.22 \pm 0.04$ & $0.25 \pm 0.05$ & $0.27 \pm 0.06$\\
	$<L_{sats}/L_{MET}>$ & $2.01 \pm 1.4$ & $2.29 \pm 1.59$ & $2.51 \pm 1.78$\\
	\enddata
	\tablecomments{Average values for our Set B runs at $t=10^8$ years. $M_{ring}$ is the rings' mass, $M_{MET}$ is the total mass of Mimas, Enceladus and Tethys, $L_{ring}$ is the rings' angular momentum, $M_{sats}$ is the total mass of the satellites in a given Run, $M_{ring,0}$ is the rings' initial mass, $L_{sats}$ is the angular momentum of the satellites in a given Run, and $L_{MET}$ is the total angular momentum of Mimas, Enceladus and Tethys.}
\end{deluxetable}

\subsection{Set C: Simulations with Dione and Rhea and no satellite tides}
For a traditional Saturn tidal paramter of $Q \sim 10^4$, mid-sized moons spawned from a Roche-interior ring do not reach distances consistent with those of Dione and Rhea.  We assume Dione and Rhea formed from a different process, which we argue is the most straightforward way to explain the much larger total mass of rock in these moons compared to that in Mimas, Enceladus, and Tethys. If Dione and Rhea were present as the inner mid-sized moons (or their progenitors) were spawned from the rings, the inner moons would have encountered mean-motion resonances with the outer satellites as the spawned moons recoiled outward due to ring interactions. For example, Dione's 2:1 MMR currently lies near $\SI{4.08}{R_{\saturn}}$ and Rhea's 3:1 MMR currently lies around $\SI{4.35}{R_{\saturn}}$, positions that a Tethys-analog would need to cross to reach Tethys' current orbit at $\SI{5.06}{R_{\saturn}}$.

\begin{figure*}[!h]
	\begin{center}
		\includegraphics[width=18cm]{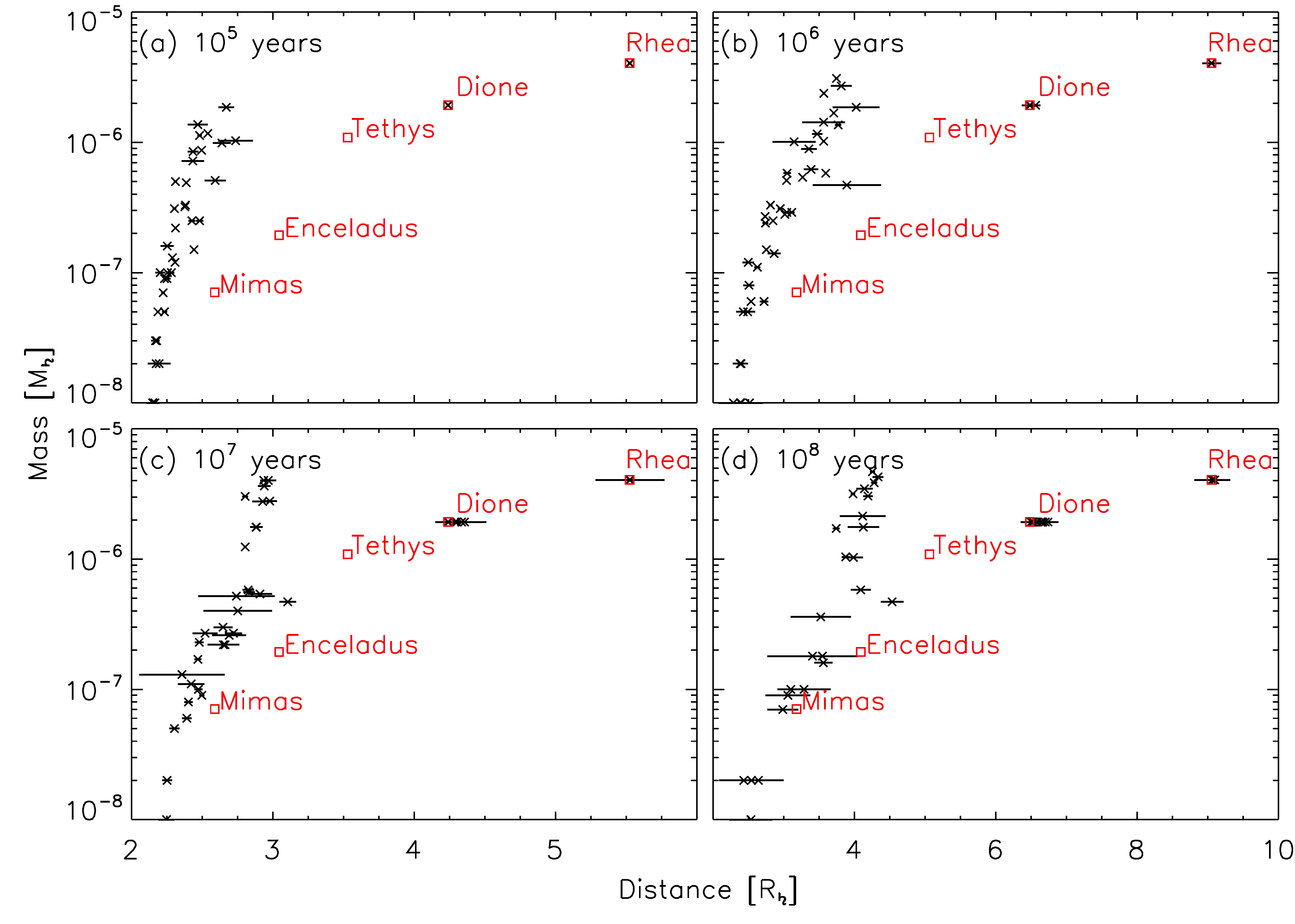}
		\caption{Distribution of satellites in our Set C simulations with Dione and Rhea at evolution times of $10^5$, $10^6$, $10^7$ and $\SI{e8}{years}$. The red squares represent the current Mimas, Enceladus, Tethys, Dione and Rhea. Horizontal lines show pericenter and apocenter of the each satellite.}
		\label{figSatDistr_DR}
	\end{center}
\end{figure*}

In the Set C runs, we include Dione and Rhea with their current masses and positions with $\mathcal{A}=0$ (no satellite tides); results are shown in Table \ref{table_setc_results} and Figure \ref{figSatDistr_DR}. 

\begin{deluxetable}{l c c c}
	\tabletypesize{\footnotesize}
	\tablewidth{0pt}
	\tablecolumns{4}
	\tablecaption{Set C data at $t=10^8$ years. \label{table_setc_results}}
	\tablehead{ & \colhead{$R_p=1.5R_{\saturn}$} & \colhead{$R_p=1.4R_{\saturn}$} & \colhead{$R_p=1.3R_{\saturn}$}}
	\startdata
	$<M_{ring}/M_{MET}>$  & $0.28 \pm 0.01$ & $0.36 \pm 0.01$ & $0.45 \pm 0.01$\\
	$<L_{ring}/L_{MET}>$  & $0.17 \pm 0.01$ & $0.22 \pm 0.01$ & $0.27 \pm 0.01$\\
	$<M_{sats}/M_{MET}>$ & $1.88 \pm 1.0$ & $2.02 \pm 1.05$ & $2.15 \pm 1.16$\\
	$<M_{sats}/M_{ring,0}>$ & $0.2 \pm 0.01$ & $0.22 \pm 0.01$ & $0.23 \pm 0.01$\\
	$<L_{sats}/L_{MET}>$ & $1.72 \pm 0.94$ & $1.87 \pm 1.01$ & $1.99\pm 1.1$\\
	$<\Delta L_{DR}/L_{MET}>$ & $0.02 \pm 0.01$ & $0.01 \pm 0.01$ & $0.01 \pm 0.01$\\
	\enddata
	\tablecomments{Average values for our Set C runs at $t=10^8$ years. $M_{ring}$ is the rings' mass, $M_{MET}$ is the total mass of Mimas, Enceladus and Tethys, $L_{ring}$ is the rings' angular momentum, $M_{sats}$ is the total mass of the spawned satellites in a given Run (excluding Dione and Rhea), $M_{ring,0}$ is the rings' initial mass, $L_{sats}$ is the angular momentum of the spawned satellites in a given Run (excluding Dione and Rhea), $L_{MET}$ is the total angular momentum of Mimas, Enceladus and Tethys, and $<\Delta L_{DR}>/L_{MET}$ is the variation of angular momentum of Dione and Rhea.}
\end{deluxetable}

The inclusion of Dione and Rhea does not dramatically alter the distribution of spawned satellites, and most results here are similar to those in Set A. The average number of satellites per run at $10^8$ years is $2.08 \pm 0.3$ (not including Dione and Rhea). No spawned moon collides with either Dione or Rhea in any of the set C runs.  However some of the spawned satellites do become captured into MMR with Dione and Rhea, as expected. These resonant configuration can be transient or still present at $10^8$ years. This has two effects. First, the inner satellites transfer some angular momentum to Dione and Rhea, resulting in Tethys-analogs that have somewhat smaller semi-major axes at $10^8$ years compared to Sets A and B.  Second, in some cases, Dione and Rhea experience substantial eccentricity growth, a result of MMRs with inner moons in the absence of tidal dissipation in the satellites. At $10^8$ years, Dione and Rhea have mean eccentricities of $0.02 \pm 0.01$ and $0.01 \pm 0.01$, respectively.

\subsection{Set D: Simulations with Dione, Rhea and satellite tides}
\begin{figure*}[!h]
	\begin{center}
		\includegraphics[width=18cm]{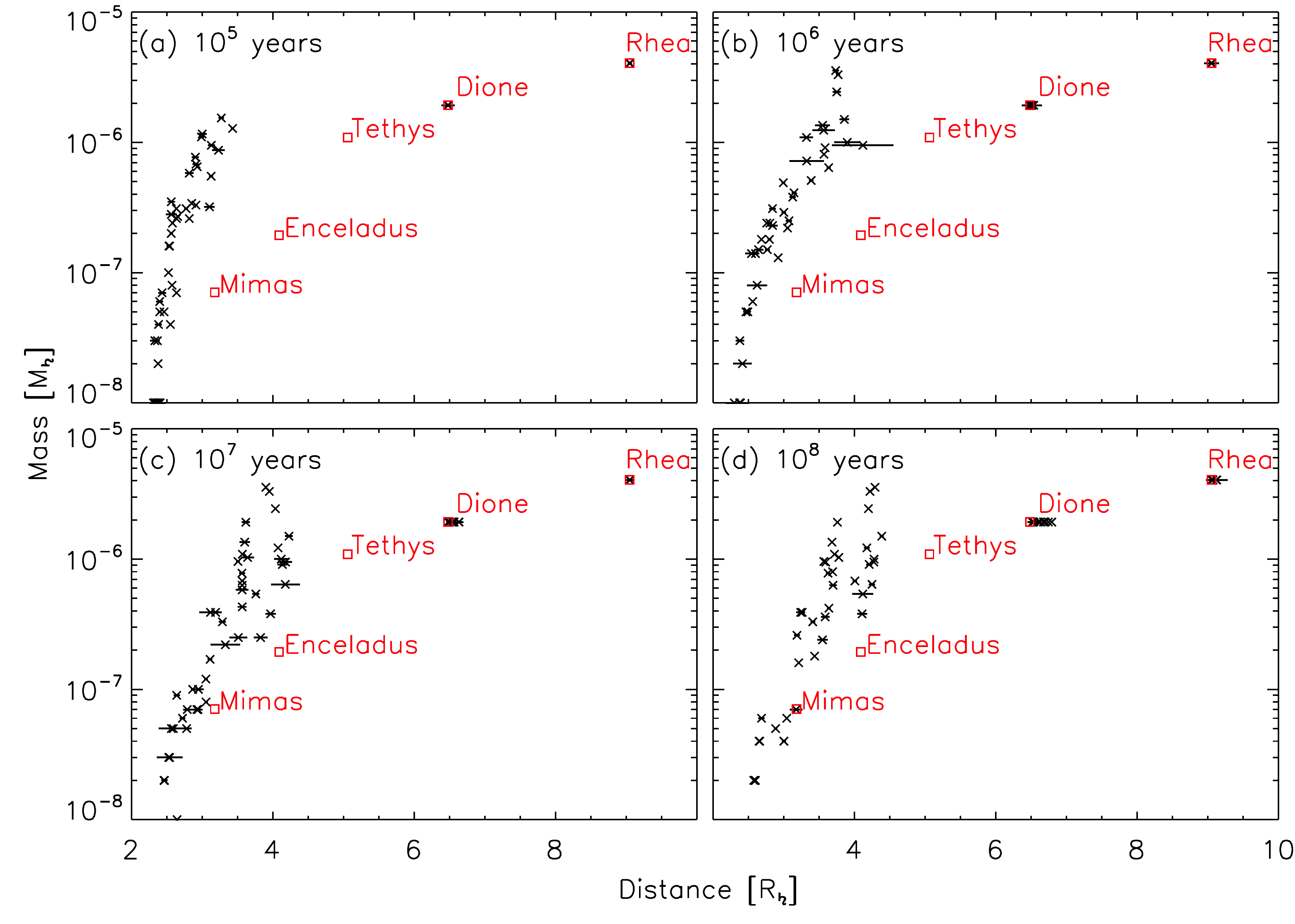}
		\caption{Distribution of satellites in our Set D simulations with Dione and Rhea, including satellite tides with $\mathcal{A}=1000$, at evolution times of $10^5$, $10^6$, $10^7$ and $\SI{e8}{years}$. The red squares represent the current Mimas, Enceladus, Tethys, Dione and Rhea. Horizontal lines represent the pericenter and apocenter of each satellite. Tidal dissipation in the satellites efficiently damps their eccentricities.}
		\label{figSatDistr_DR_SatTides}
	\end{center}
\end{figure*}

Set D simulations begin with Dione and Rhea in their current positions, and include tidal dissipation in the satellites with $\mathcal{A}=10^3$. Table \ref{table_setd_results} and Figure \ref{figSatDistr_DR_SatTides} show results at $t=\SI{e8}{years}$.  Set D simulations produce $3.4 \pm 0.5$ spawned satellites at $\SI{e8}{years}$ (excluding Dione and Rhea). Compared with the case without satellites tides (Set C), we find that less mass from the ring is placed into satellites by $10^8$ yr, and the spawned satellites have lower total angular momentum. In Set D, tidal damping of eccentricities keeps the eccentricities of Dione and Rhea lower (of order $10^{-3}$ on average), and maintains some of the spawned moons in MMRs with Dione and Rhea. We find that the largest spawned moon is captured in Dione's 2:1 MMR in 9 of the 12 runs from Set D, with the resonance configuration still present at $10^8$ years. As inner spawned moons transfer more of the ring's angular momentum to Dione and Rhea, the spawned moon orbits do not expand as far as in the case without satellite tides, resulting in greater confinement of the rings and somewhat less total mass incorporated into the spawned moons.

\begin{deluxetable}{l c c c}
	\tabletypesize{\footnotesize}
	\tablewidth{0pt}
	\tablecolumns{4}
	\tablecaption{Set D data at $t=10^8$ years. \label{table_setd_results}}
	\tablehead{ & \colhead{$R_p=1.5R_{\saturn}$} & \colhead{$R_p=1.4R_{\saturn}$} & \colhead{$R_p=1.3R_{\saturn}$}}
	\startdata
	$<M_{ring}/M_{MET}>$  & $0.26 \pm 0.05$ & $0.34 \pm 0.03$ & $0.4 \pm 0.05$\\
	$<L_{ring}/L_{MET}>$  & $0.16 \pm 0.03$ & $0.21 \pm 0.02$ & $0.24 \pm 0.03$\\
	$<M_{sats}/M_{MET}>$ & $1.66 \pm 0.94$ & $1.84 \pm 1.01$ & $2.04 \pm 1.12$\\
	$<M_{sats}/M_{ring,0}>$ & $0.18 \pm 0.01$ & $0.19 \pm 0.01$ & $0.22 \pm 0.01$\\
	$<L_{sats}/L_{MET}>$ & $1.49 \pm 0.86$ & $1.67 \pm 0.95$ & $1.85 \pm 1.05$\\
	$<\Delta L_{DR}/L_{MET}>$ & $0.03 \pm 0.01$ & $0.02 \pm 0.01$ & $0.02 \pm 0.02$\\
	\enddata
	\tablecomments{Average values for our Set D runs at $t=10^8$ years. $M_{ring}$ is the rings' mass, $M_{MET}$ is the total mass of Mimas, Enceladus and Tethys, $L_{ring}$ is the rings' angular momentum, $M_{sats}$ is the total mass of the satellites in a given Run (excluding Dione and Rhea), $M_{ring,0}$ is the rings' initial mass, $L_{sats}$ is the angular momentum of the satellites in a given Run (excluding Dione and Rhea), $L_{MET}$ is the total angular momentum of Mimas, Enceladus and Tethys, and $<\Delta L_{DR}>/L_{MET}$ is the variation of angular momentum of Dione and Rhea.}
\end{deluxetable}

Figure \ref{figSatDistr_DR_SatTides} shows the obtained distribution of satellites, at different times of evolution. As with the case without Dione and Rhea, the inclusion of tidal dissipation in the satellites efficiently damps the eccentricities of the growing moons. As a consequence, the radial ``feeding'' zone of each satellite is narrower as they remain on quasi-circular orbits, which allows on average a larger number of satellites to survive, in particular Mimas progenitors lying close to the rings.

\section{Simulations over $\SI{e9}{years}$.} \label{Section_Simulations1d9}

Each of the 48 simulations described above required about 1.5 months of computational time to track the system evolution for $10^8$ yr.  It is clear that these systems would continue to evolve over longer timescales, given the mass and angular momentum still in the rings at $10^8$ yr time and expected further tidal evolution.  We wish to track the system's evolution over an order-of-magnitude longer $10^9$-yr timescale to determine whether the resulting distribution of spawned moons will approach that of the current inner moons.  

Billion-year simulations of circumplanetary material require a modified numerical approach to be computationally feasible. As such we develop an accelerated version of our model to approximate the evolution in this regime, wherein we multiply the disk's viscosity, and the strength of tides and resonant interactions all by a factor of 10. In the accelerated code, the relative rates of viscous spreading, ring torques and tidal evolution are thus the same as in our default simulations.  However the absolute rates of these processes are an order-of-magnitude faster compared to the orbital frequency at a particular radius in the disk.  Because in general the orbital timescales are much shorter than the timescales associated with the other processes, the overall evolution of the system may be well-approximated by such an accelerated treatment.  However the accelerated approach may miss some aspects of the dynamics; e.g., it is possible that in the accelerated code an object's orbit may evolve to quickly to become captured into a MMR when that same object would have been captured with a slower orbital evolution.   
 
\subsection{Test of the accelerated code.}
\begin{figure}[!h]
	\begin{center}
		\includegraphics[width=8cm]{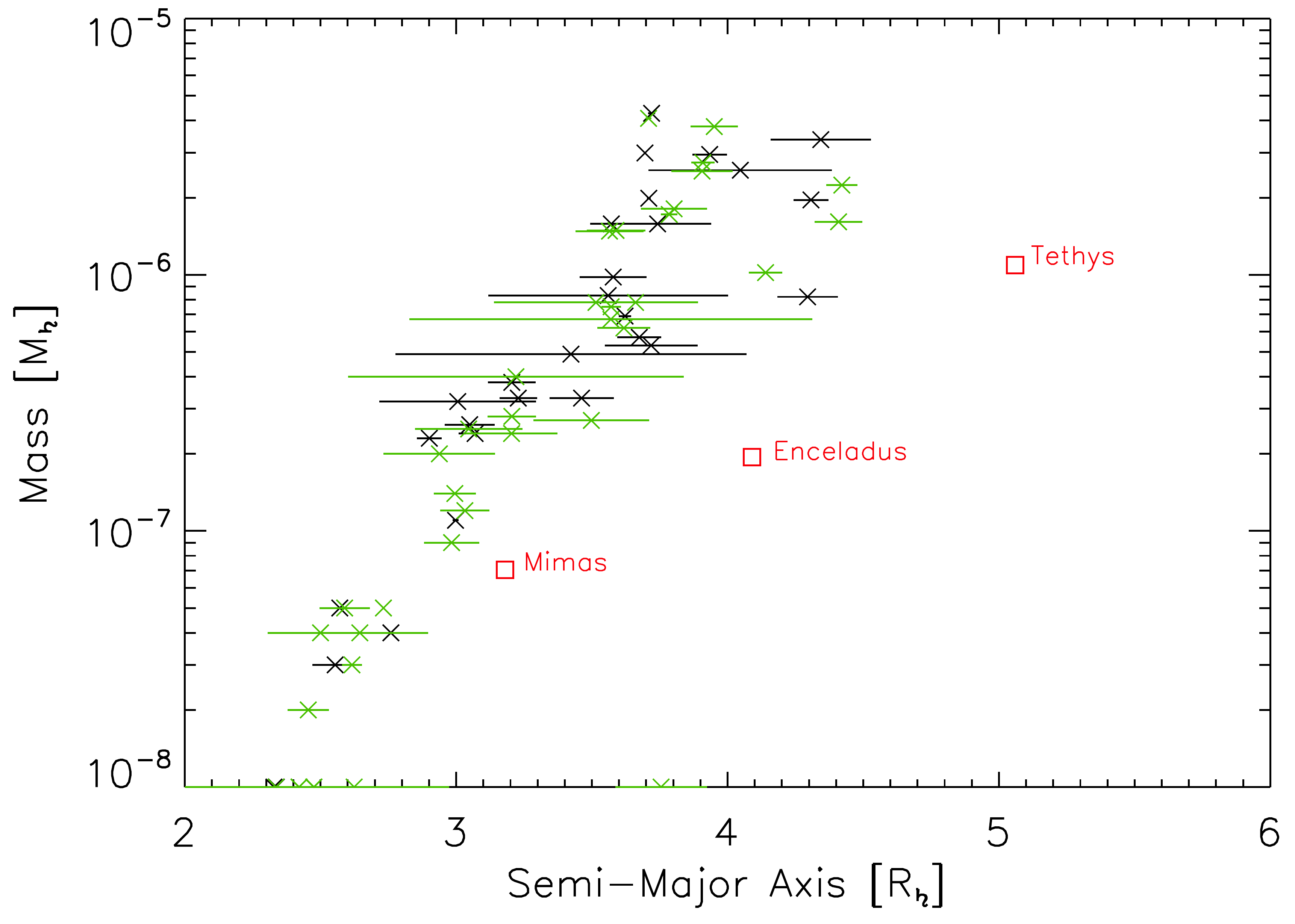}
		\caption{Distribution of satellites obtained with the standard code at $\SI{e7}{years}$ (black crosses) and with the accelerated code at $\SI{e6}{years}$ (green crosses). Horizontal lines represent the pericenter and apocenter of the object.}
		\label{figAccelTest}
	\end{center}
\end{figure}

We ran the 12 Set A simulations for $\SI{e6}{years}$ with the accelerated code and compared the obtained distribution of satellites with that obtained over $\SI{e7}{years}$ with the standard code. The results are shown in Figure \ref{figAccelTest}. The average number of satellites per run is $3.17 \pm 1.03$ with the accelerated code, compared to $2.42 \pm 0.79$ with the standard code. This is mainly due to a greater number of small satellites in the accelerated code. If we consider only satellites with a mass $\ge \SI{2e-8}{M_{\saturn}}$, thereby removing the stochasticity of newly spawned moonlets, then the average number of satellites per run is $2.25 \pm 0.96$ with the standard code, and $2.58 \pm 0.79$ with the accelerated code. Average eccentricity of satellites are $0.05 \pm 0.066$ for the accelerated runs, and $0.05 \pm 0.058$ for the normal ones. Thus the overall distribution of spawned satellites produced by our accelerated code is similar to those obtained with the normal code.

\subsection{Contraction of the planet}
For our initial runs we have assumed that the planet radius remained constant over $\SI{e8}{years}$, with a value of 1.3, 1.4, or 1.5$R_{\saturn}$. For our billion-year runs, we use as our starting condition the outputs of Runs 5 to 12 in Sets A through D.  Thus the initial planetary radius is either 1.3 or $\SI{1.4}{R_{\saturn}}$.  We then include the contraction of the planet and increase its spin (which moves the synchronous orbit inward) in the $10^9$-year runs. This is done by first estimating the initial age $t_{ev}$ of the planet given its radius (i.e. inverting Figure \ref{figSaturnContraction}), and then by computing the new planetary radius at time $t$ given by $R(t_{ev}+t)$. To save computational time, we do not update the radius at every timestep. We find that updating the radius every $\sim \SI{e4}{years}$ gives a good approximation of our analytical derivation from Section \ref{Subsection_SaturnRadius}.

\subsection{Results}
\begin{figure}[!h]
	\begin{center}
		\includegraphics[width=8.5cm]{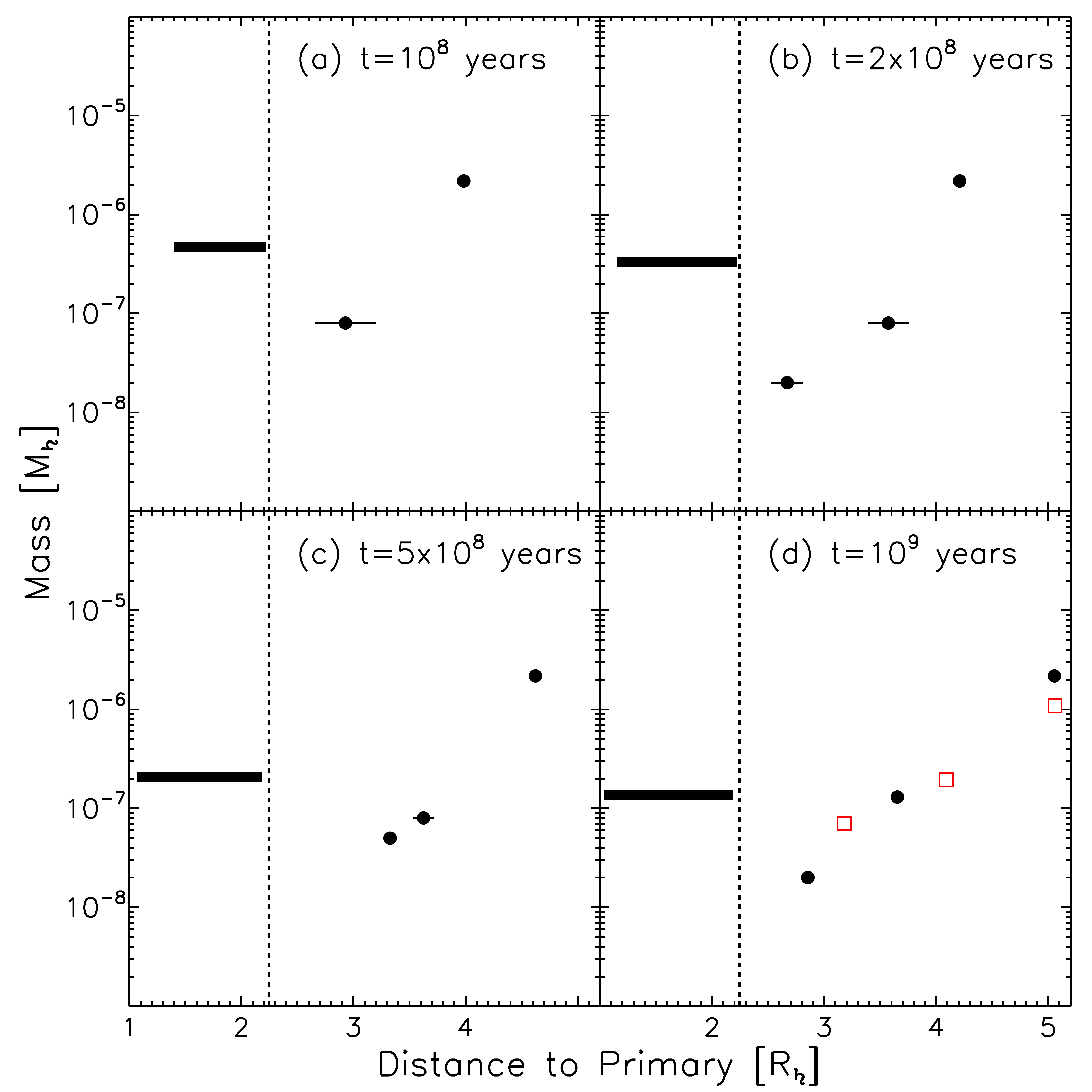}
		\includegraphics[width=8.5cm]{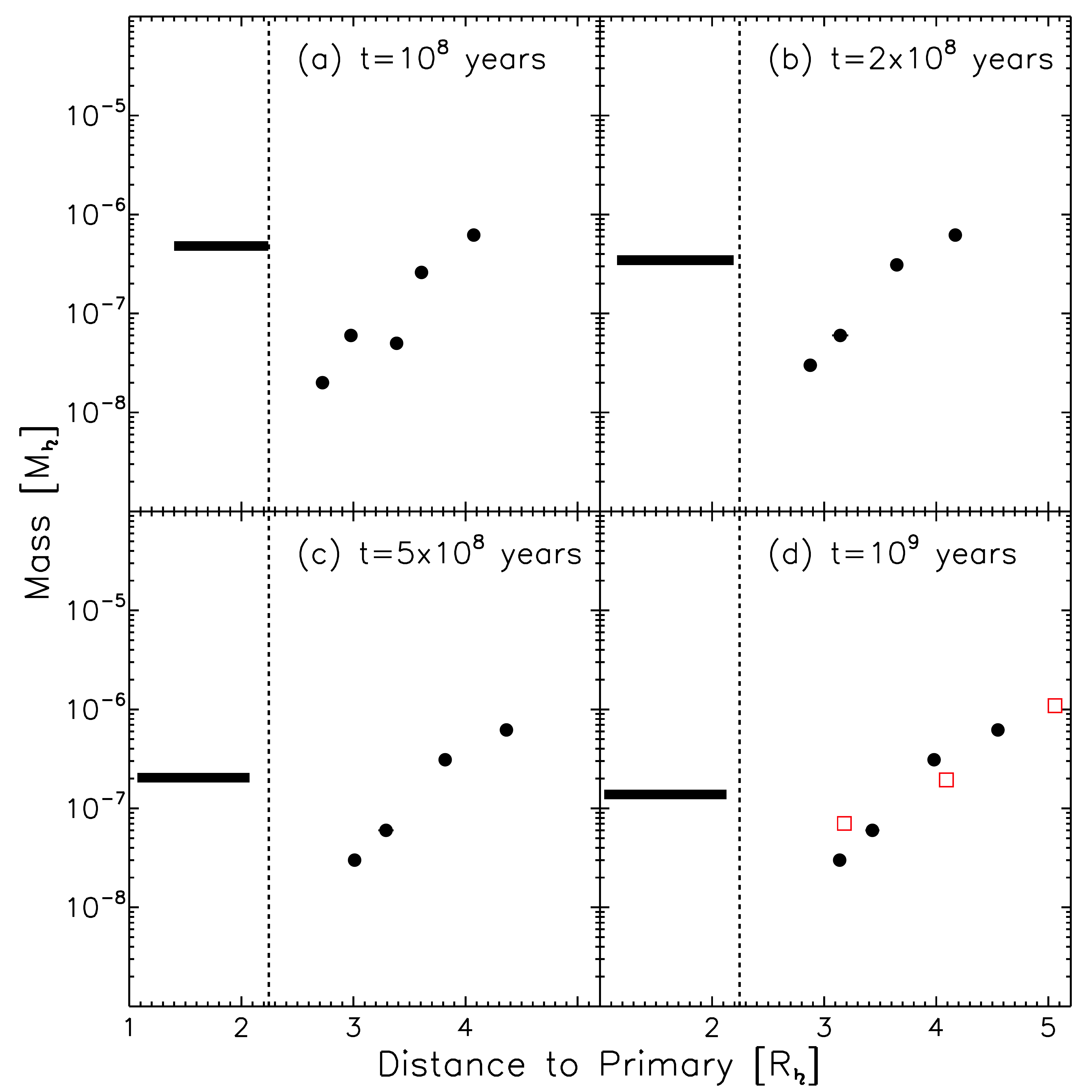}
		\caption{Snapshots of the system in Run 6A (left) and 5B (right) at different times of evolution. The vertical dashed line at $\approx 2.24R_{\saturn}$ is the Roche limit. The thick black horizontal line is the Roche-interior ring, whose inner edge moves inward as the planet's contracts from its initial radius of $\SI{1.3}{R_{\saturn}}$. The black dots represent the satellites formed from the disk, with the thin horizontal lines representing their pericenter and apocenter. The red squares show the current Mimas, Enceladus and Tethys.}
		\label{figSimuPannels1d9}
	\end{center}
\end{figure}

Figure \ref{figSimuPannels1d9} shows the later evolution of the system from Run 6A on the left panel and for Run 5B on the right panel. In both cases the largest satellite spawned in the first $10^8$ years does not accrete additional mass but keeps evolving away due to tides and capture of inner moons in MMRs. The second largest satellite has continued growing, and reaches its final mass after a few $10^8$ years. The formation of a Mimas-equivalent satellite is not complete in Run 6A. In Run 5B, there are 2 moons around the position of current Mimas, and they will likely merge over longer timescales. This is also seen in the other runs where we get good Enceladus and Tethys equivalent satellites. On the other hand, cases where we form a good Mimas analog at $10^9$ years have either too many moons outside Mimas, or 2 moons that are much more massive than Enceladus and Tethys.

In Run6A, the rings still contain $\sim \SI{1.4e-7}{M_{\saturn}}$ and should be able to provide enough material to complete the formation of a Mimas-type satellite over longer timescales. A similar conclusion can be made for the other runs that have good Enceladus and Tethys equivalents at $10^9$ years. This however implies that Mimas may be at least a billion year younger than Tethys. A similar point was made by \cite{charnoz11} who claimed that ``a Mimas-like satellite could be about $1-1.5$ Gy younger than a Rhea-like satellite''.

\begin{figure*}[!h]
	\begin{center}
		\includegraphics[width=18cm]{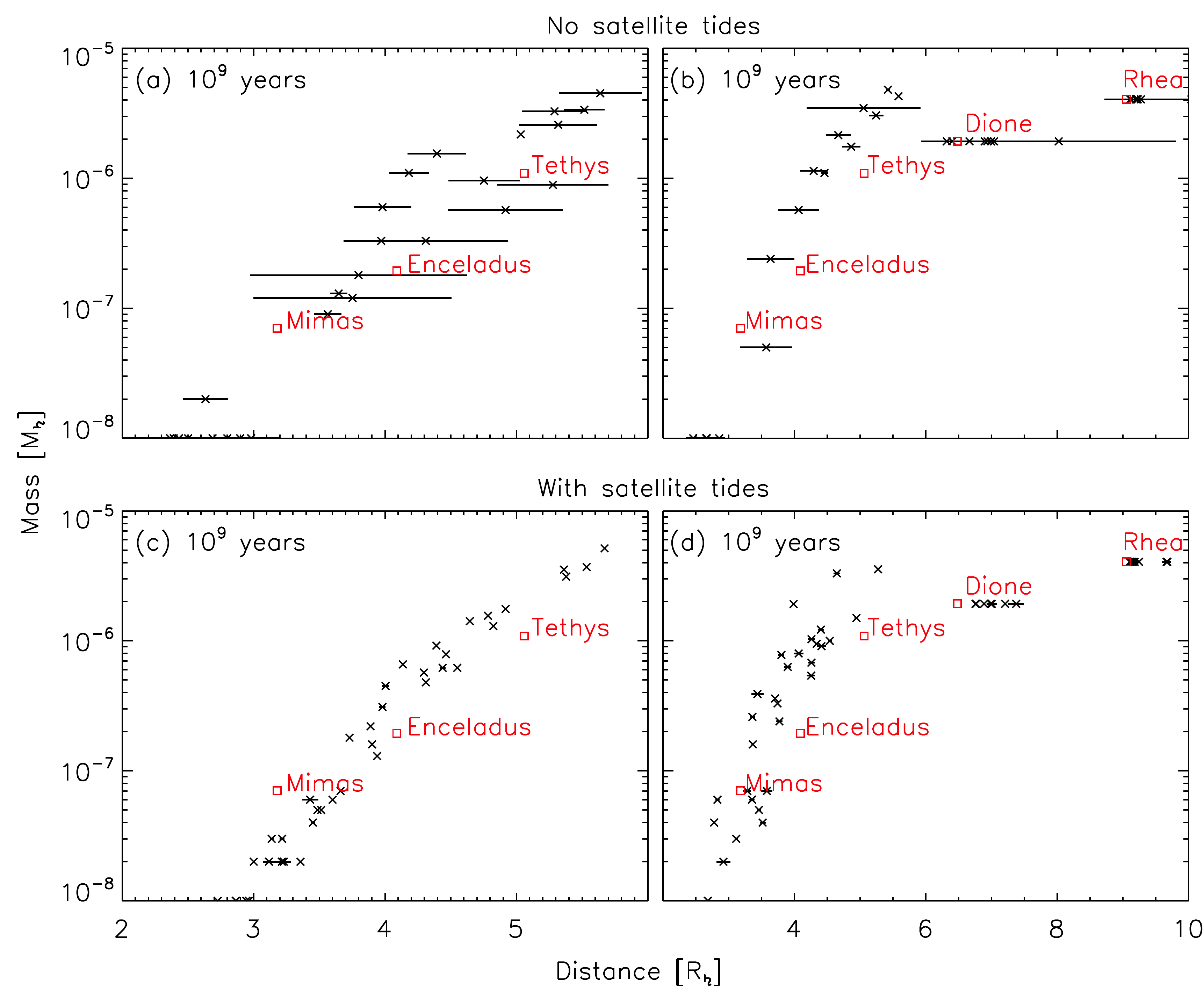}
		\caption{Distribution of satellites at $\SI{e9}{years}$. a) Without Dione and Rhea nor satellites tides. b) With Dione and Rhea but without satellite tides. c) Without Dione and Rhea but including satellite tides. d) With Dione and Rhea and including satellite tides.}
		\label{figSatDistr1d9}
	\end{center}
\end{figure*}

Figure \ref{figSatDistr1d9} shows the distribution of satellites at $\SI{e9}{years}$ for all cases considered previously (with/without Dione and Rhea, with/without satellite tides). For the cases without Dione and Rhea (panels a and c), the distribution of satellites obtained in our various runs agrees reasonably well with the masses and positions of current Mimas, Enceladus and Tethys. The runs that include tidal dissipation in the satellites (panels c and d) are better at producing Mimas equivalents. This is due to the fact that tidal dissipation keeps the eccentricity of the outer satellites small, such that their pericenter lies further from the edge of the rings, preventing them from ``sweeping'' this area. The average number of satellites per run at $10^9$ years is $3.25 \pm 0.7$ for Set A, $4.87 \pm 0.64$ for Set B, $2.12 \pm 1.13$ for Set C and $3.75 \pm 0.71$ for Set D. As expected, the number of satellites per runs is larger in cases that include tidal dissipation in the satellites as the latter keeps eccentricities low and limits the feeding zone of a given satellite, allowing more satellites to coexist.

For the set of runs that include Dione and Rhea (panels b and d), the match to Mimas, Enceladus and Tethys is best when we include satellite tides (Set D). Through capture into MMR, the inner satellites have transfered some of their angular momentum to Dione and/or Rhea, and have thus experienced weaker orbital migration. On some cases, Dione and Rhea have migrated away significantly. As for the runs without Dione and Rhea, the inclusion of tidal dissipation into the satellites allows the survival of Mimas like satellites.

Table \ref{table_1d9_results} lists average quantities for the rings and produced satellites at $10^9$ years across our 4 Sets of simulations. The average mass of the rings at $\SI{e9}{years}$ is $\num{1.5e-7} \pm \SI{2e-8}{M_{\saturn}}$ across all our runs. Despite a factor of six variation in the initial mass, all rings have roughly the same mass at the end of our simulations, in good agreement with the expected asymptotic evolution of the ring mass \cite{salmon10}. The average ring mass at $\SI{e9}{years}$ is only slightly larger than the mass of Mimas, in good agreement with current estimates of the mass of Saturn's rings \citep{esposito83,salmon10}.

\begin{deluxetable}{l c c c c}
	\tabletypesize{\footnotesize}
	\tablewidth{0pt}
	\tablecolumns{5}
	\tablecaption{Data for all 4 Sets at $t=10^9$ years. \label{table_1d9_results}}
	\tablehead{ & \colhead{Set A} & \colhead{Set B} & \colhead{Set C} & \colhead{Set D}}
	\startdata
	$<M_{ring}/M_{MET}>$      & $0.12 \pm 0.02$ & $0.12 \pm 0.01$ & $0.12 \pm 0.02$ & $0.11 \pm 0.02$\\
	$<L_{ring}/L_{MET}>$      & $0.07 \pm 0.01$ & $0.07 \pm 0.01$ & $0.07 \pm 0.01$ & $0.06 \pm 0.01$\\
	$<M_{sats}/M_{MET}>$      & $2.12 \pm 1.02$ & $2.62 \pm 1.66$ & $2.13 \pm 1.02$ & $1.95 \pm 1.01$\\
	$<M_{sats}/M_{ring,0}>$   & $0.22 \pm 0.01$ & $0.25 \pm 0.05$ & $0.22 \pm 0.01$ & $0.2  \pm 0.01$\\
	$<L_{sats}/L_{MET}>$      & $2.18 \pm 1.1$  & $2.67 \pm 1.78$ & $2.2  \pm 1.13$ & $1.87 \pm 1.03$\\
	$<\Delta L_{DR}/L_{MET}>$ & N/A             & N/A             & $0.09 \pm 0.1 $ & $0.11 \pm 0.08$\\
	\enddata
	\tablecomments{Average values at $10^9$ for the 4 sets of simulations. $M_{ring}$ is the rings' mass, $M_{MET}$ is the total mass of Mimas, Enceladus and Tethys, $L_{ring}$ is the rings' angular momentum, $M_{sats}$ is the total mass of the spawned satellites in a given Run (excluding Dione and Rhea), $M_{ring,0}$ is the rings' initial mass, $L_{sats}$ is the angular momentum of the spawned satellites in a given Run (excluding Dione and Rhea), $L_{MET}$ is the total angular momentum of Mimas, Enceladus and Tethys, and $<\Delta L_{DR}/L_{MET}>$ is the variation of angular momentum of Dione and Rhea.}
\end{deluxetable}

\begin{figure*}[!h]
	\begin{center}
		\includegraphics[width=9cm]{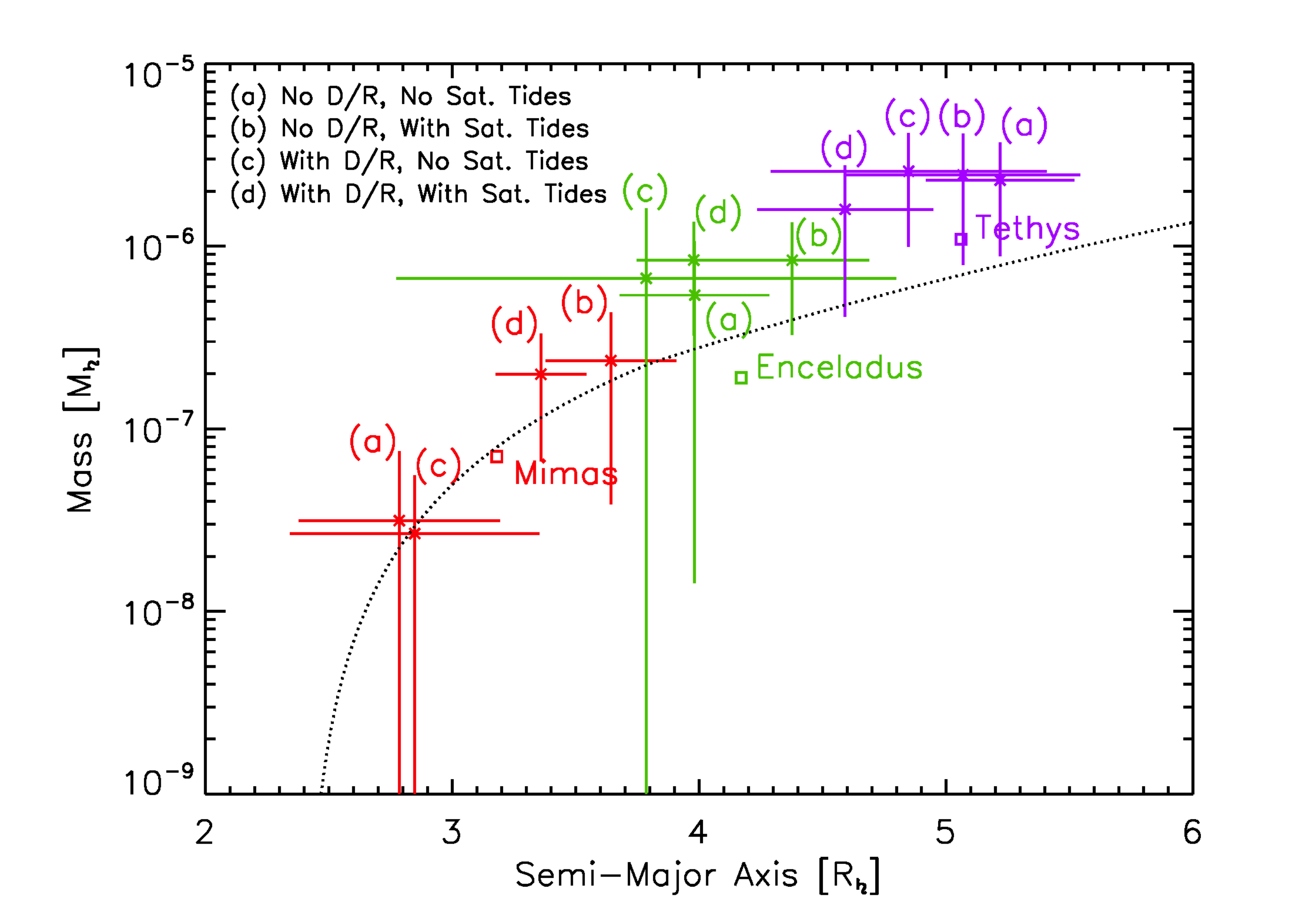}
		\caption{Average mass and semi-major axis of satellites formed in our 4 sets of simulations. For the innermost satellite, we use the total mass of the 3rd most massive satellite plus any other moonlets inside its orbit, and compute a mass-weighted semi-major axis. The dotted line is the analytical prediction from \cite{crida12} (their equation 25S). Without satellite tides (Sets A and C) the innermost satellite is less massive than Mimas and inside its current orbit, while with strong satellite tides (Sets B and D), it is more massive and outside. Moderate satellite tides would likely produce a better fit to current Mimas. Our Enceladus-equivalent is always significantly more massive than the current one, consistent with later mass loss \citep[e.g.][]{hansen08}}
		\label{fig_SatRange}
	\end{center}
\end{figure*}

Figure \ref{fig_SatRange} shows the range of satellites formed across all our simulations. We plot the average mass and semi-major axis of our Tethys-, Enceladus-, and Mimas-analogs. For the latter, we sum the masses of the 3rd largest bodies with any other moonlets present inside its orbit, as they will likely collide and merge on longer timescales, and we compute a mass-weighted mean for the semi-major axis. For our outermost satellite, the mass range is similar in sets A, B and C and a little lower for set D. For the outermost satellite, sets not including tidal dissipation in the satellites (A and C) have consistently larger average semi-major axes than the corresponding set that includes them (B and D). Our Enceladus-analogs are systematically larger than current Enceladus, which is consistent with later mass loss from this body \citep[e.g.][]{hansen08}.

For the innermost satellite there is a clear separation between sets that do not include satellite tides, which produce smaller and closer moons, and sets that include strong satellite tides, which produce larger and more distant satellites. While we have in this study explored extreme cases (no satellite tides or very strong ones), this suggests that moderate satellite tides could produce a innermost satellite more similar to Mimas.

We find that many of our produced satellite systems show resonant configurations at $10^9$ years: 3 runs in Set A, 5 runs in Set B, 3 runs in Set C and 6 runs in Set D. For Set C and D, a resonant configuration between Dione and Rhea occurs in 2 and 5 runs respectively. In Set D, the largest moon formed from the rings (i.e. our Tethys equivalent) is captured in a Dione's 2:1 MMR in 5 runs. Despite efficient capture into MMRs in all our cases, survival of the system of Saturn's moons through their formation appears likely over 1 billion years. Some tidal dissipation into the growing satellites seems a prerequisite to allow the formation of the innermost satellites, in particular Mimas.

\section{External delivery of rock to Saturn's inner moons}
Mimas, Enceladus and Tethys as a group contain between $6 \times 10^{19}$ to $10^{20}$ kg rock, about $8$ to $15\%$ of their combined current masses (Table 1). Enceladus is currently about half rock, although its initial proportion of rock may have been much lower if it has lost ice over its history at a rate comparable to that occurring currently as a result of its endogenic acitivy.   Clearly if Saturn's inner moons (or their progenitors) were spawned from an essentially pure ice ring as we consider here, the moons would have initially been nearly pure ice as well.   Here we consider how external bombardment onto these moons might have altered their initial compositions and possibly supplied the rock in these objects.

\subsection{A late heavy bombardment in the outer solar system}

In the so-called “Nice” model for the origin of the dynamical structure of the outer solar system \citep{tsiganis05}, the giant planets are initially in a compact orbital configuration. Their orbits slowly migrate and diverge due to dynamical interactions with a planetesimal disk of initial mass $M_{disk}$.  When Jupiter and Saturn cross their mutual 2:1 MMR, their orbital eccentricities increase.  This leads to a period of dynamical instability, during which Uranus and Neptune (and perhaps a fifth outer planet as well; \cite{nesvorny12}) are scattered outward to their current positions and planetesimals are scattered across the solar system.  The scattered planetesimals become a population of impactors, which could, e.g., account for the spike in impact rate believed by many to be necessary to explain the formation of the large lunar impact basins at $\sim 3.9$ Gyr during a so-called “late heavy bombardment” on the Moon \citep{tera74,cohen00,stoffler01,kring02,gomes05,levison11}. The total mass of scattered planetesimals scales roughly with $M_{disk}$.  The disk must be massive enough to decrease the eccentricities of Uranus and Neptune to their current values through friction effects, but not so massive as to cause excessive migration.   Forming a planetary system like ours appears most likely for $20M_{\oplus}\le M_{disk} \le 50M_{\oplus}$ \citep{batygin10,nesvorny12}.  For $M_{disk} = 35M_{\oplus}$, $\sim 10^{19}$ kg of planetesimals are scattered onto the Moon, consistent with that needed to explain the lunar LHB \citep{gomes05}. If the instability was responsible for the lunar LHB, the timing of the event is constrained to occur some 700 Myr after the origin of the solar system.

A Nice-like instability consistent with our specific solar system structure may be a relatively improbable case among the broad range of possible outcomes \citep{nesvorny12}.  However the Nice model remains the most detailed dynamical history available for the early outer solar system, and it has been remarkably successful in explaining a variety of solar system features, including the giant planet eccentricities \citep{tsiganis05}, the structure of the Kuiper Belt \citep{levison08}, capture of Jupiter’s Trojans \citep{morbidelli05}, capture of the irregular satellites \citep{nesvorny07}, and the Ganymede/Callisto dichotomy \citep{barr10}.  While we consider predictions of the specific Nice model as a guide, the existence of an enhanced bombardment period in the outer solar system probably applies more generally.  It has long been recognized that the structure of the Kuiper Belt requires that Neptune migrated outward via planetesimal scattering \citep{malhotra95}, and that this implies both an initially more compact giant planet configuration and a planetesimal disk containing between 10 and 100$M_{\oplus}$ \citep{fernandez84,hahn99}.  The interaction of the giant planets with such a massive disk would have produced an enhanced bombardment period even if the details of the evolution differed from that of the Nice model.  

The composition of the impactors originating from the plaentesimal disk is uncertain. Jupiter trojans are thought to be captured bodies that originated in the region beyond Neptune \citep{nesvorny13}. These bodies show a diversity of densities: $\SI{0.8}{g.cm^{-3}}$ for 617 Patroclus \citep{marchis06}, and $\SI{2.5}{g.cm^{-3}}$ for 624 Hektor \citep{lacerda07}. Jupiter-family comets are also thought to originate from the Kuiper Belt \citep{fernandez80,duncan88}, and show densities lower than $\SI{0.6}{g.cm^{-3}}$ \citep{lowry08}. These low densities indicate high porosities. The rock fraction of these objects would be between about 40 and $70\%$ if they reflected a bulk solar composition of outer solar system solids \citep{simonelli89}. Observations from the Deep Impact mission imply that the rock-to-ice ratio in comet 9P/Tempel 1 is larger than one, suggesting comets are ``icy dirtballs'' rather than ``dirty iceballs'' \citep{kuppers05}. The size distribution of impactors is also uncertain, but constraints can be derived from the cratering rate on Iapetus, and the observed size distribution of comets and KBOs \citep{charnoz09}.   

\subsection{Mass of rock delivered to inner spawned moons at Saturn}

A late heavy bombardment in the outer solar system would have affected any satellite of Saturn that existed at that time. We here consider a bombardment that occurs at 700 Myr, and estimate the mass of rock that would have collided with Saturn's inner moons as a function of the assumed mass of the transneptunian disk, assuming that the LHB impactors contain between $40\%$ and $60\%$ rock by mass.   

For an initial disk mass $M_{d,0}=\SI{35}{M_{\oplus}}$, the Moon accretes an average of
$\SI[separate-uncertainty]{8.4+-0.3e18}{kg}$ \citep{gomes05}, which is $\sim \SI{4e-8}{M_{d,0}}$. \cite{levison01} found that during and LHB-type event, Callisto and Ganymede are respectively impacted 40 and 110 times more than the Moon. Using the impact probabilities from Table 1 of \cite{zahnle03}, this implies that a mass $\sim \SI{3.14e-2}{M_{d,0}}$ collides with Jupiter, while $\sim \SI{1.32e-2}{M_{d,0}}$ collides with Saturn, where these values assume the planets have their current mean radii.

Given the probability of impact with Saturn, $P_{\saturn}$, the impact probability onto a Saturnian satellite, $P_s$, can be approximated by \citep{zahnle03}:

\begin{equation}
\frac{P_s}{P_{\saturn}}\approx \frac{R_s^2}{R_{\saturn} a_s},
\label{equZahnle}
\end{equation}
where $R_s$ and $a_s$ are the satellite's physical radius and its semi-major axis. 

\begin{figure}
	\begin{center}
		\includegraphics[width=8cm]{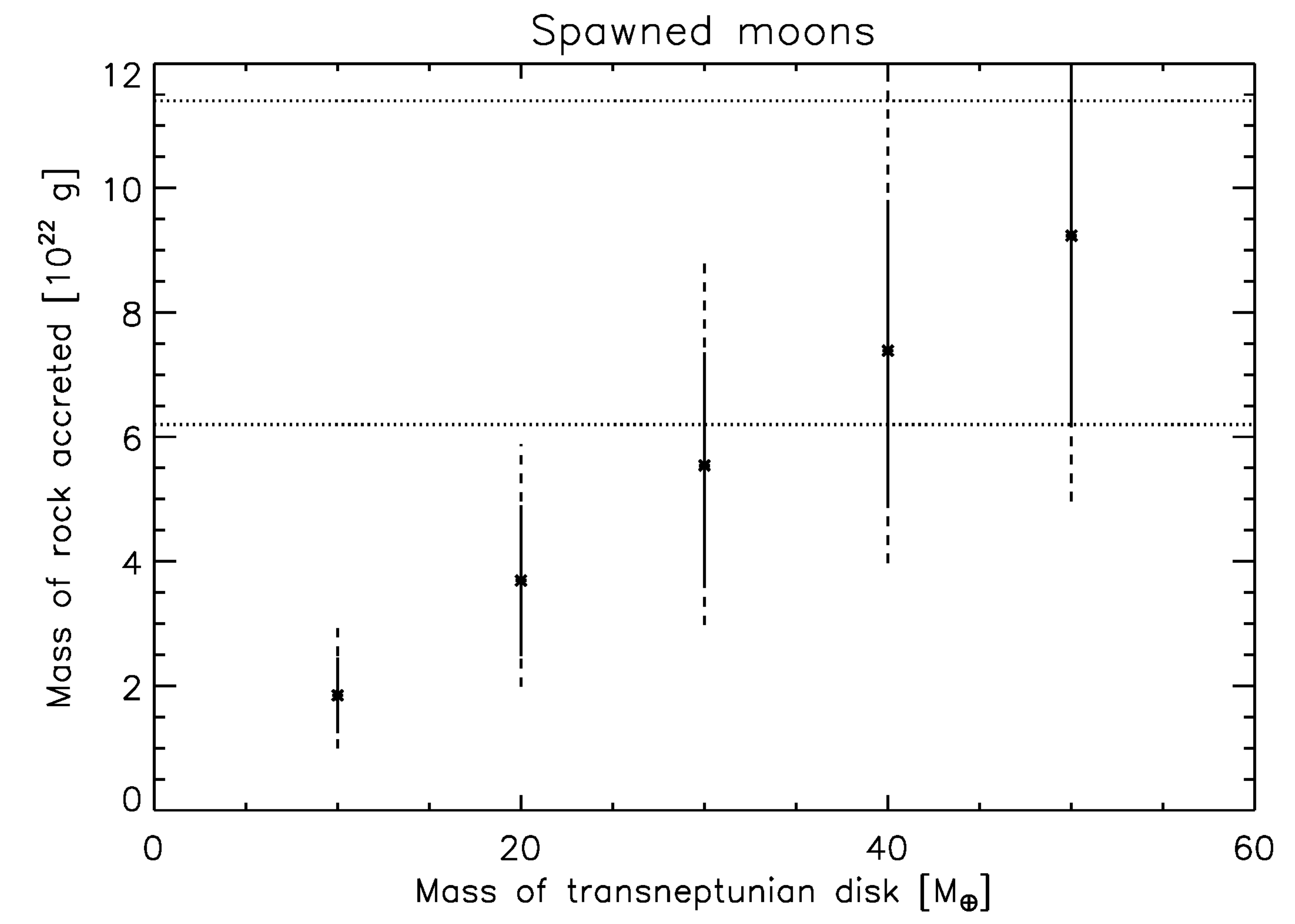}
		\includegraphics[width=8cm]{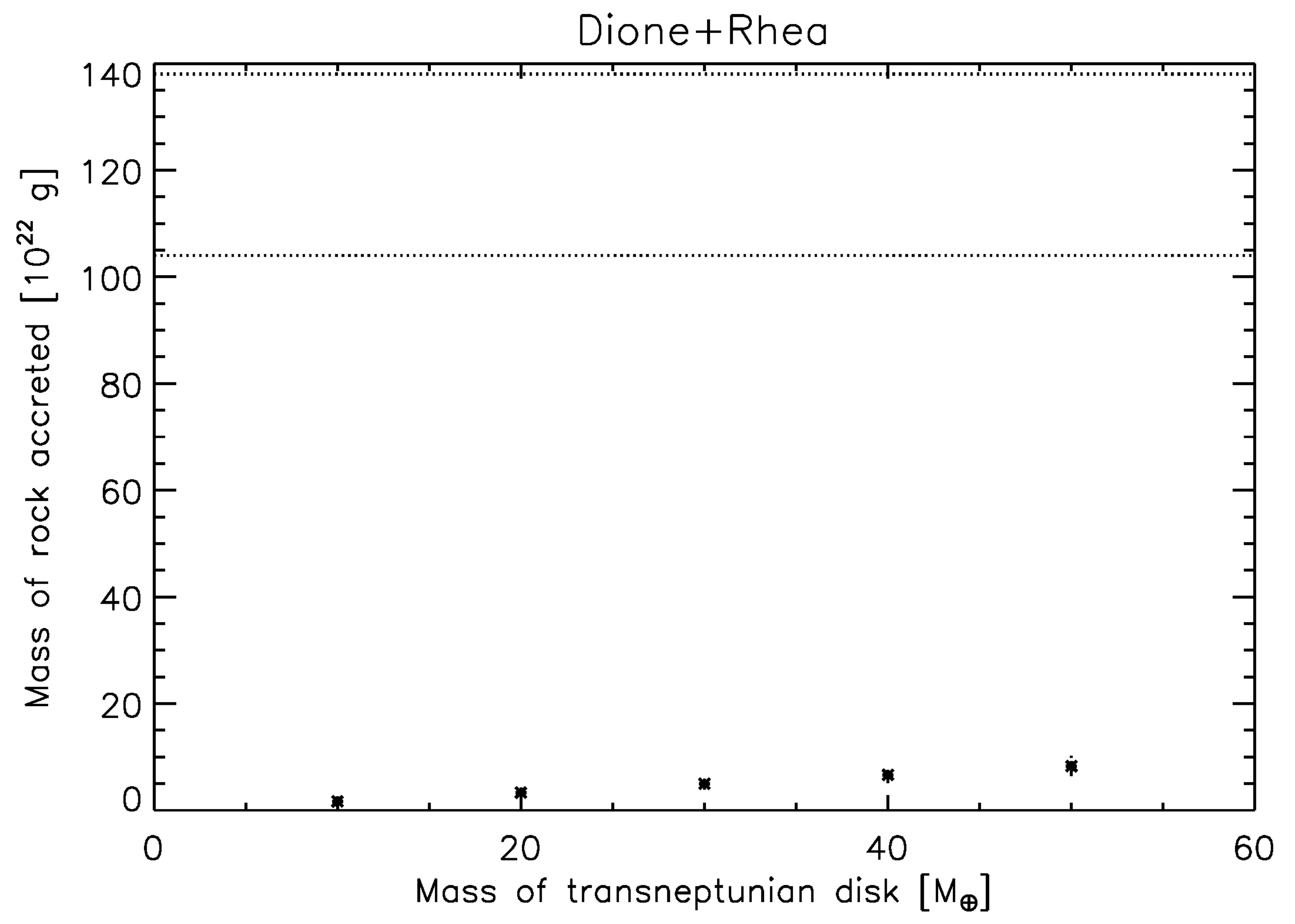}
		\caption{Average mass of rock accreted by the satellites as a function of the initial mass of the transneptunian disk. We assume that the LHB occurs at $\SI{700}{Myr}$ and that impactors contain $50\%$ of rock in mass. (left) Points show the average total mass accreted by satellites spawned from the rings in our simulations; vertical lines indicate standard deviation. The horizontal dotted lines show the range of total rock mass in Mimas, Enceladus, and Tethys today. The vertical dashes represent variations in the accreted mass for impactors containing 40 to 60\% rock. (right) Points show the mass of rock accreted by Dione and Rhea (when present) as a second group, which is much less than the total rock contained in these two moons indicated by the horizontal dashed lines.}
		\label{fig_mrock_average}
	\end{center}
\end{figure}

We apply this formula to the the distribution of moons in each of our simulations at $t = 700$ Myr to determine the probability of impact with each moon.  For the spawned moons, we calculate $R_s$ assuming a density appropriate for pure ice; for Dione and Rhea, we use their current physical radii.  We estimate the total mass of rock delivered to the satellites during the LHB by computing the total of these probabilities times $M_{d,0}$.  Figure \ref{fig_mrock_average} shows results as a function of $M_{d,0}$. The points indicate the average mass of rock (and its standard deviation) by satellites spawned from the rings (left panel), and by Dione and Rhea in runs that included them (right panel). The vertical dashed lines represent variations for impactors containing 40 to 60\% rock.

The mass of rock that impacts the satellites spawned from the rings is consistent with the total mass of rock in Mimas, Enceladus and Tethys (shown as the horizontal dashed lines in Fig. \ref{fig_mrock_average}) for $M_{d,0}>\SI{20}{M_{\oplus}}$. Thus if an LHB in the outer solar system occcurred at roughly the same time as the lunar LHB, it would have delivered a rock mass comparable to the total rock in Saturn's inner moons. This suggests that these moons were initially much more rock-poor than they are today.  In contrast, the LHB delivers a mass in rock to Dione and Rhea that is about an order-of-magnitude less than the current rock content in these moons (Figure 17, right panel), suggesting that they were already (relatively) rock-rich when they formed.  

In the above we estimate the total mass of rock from external impactors that collides with the spawned satellites as a group, but we have not calculated the rock mass that would ultimately be accreted by each satellite. Using the mean values for each equivalent satellite (Figure \ref{fig_SatRange}), we can estimate the expected rock mass to collide with each object. We find that ``Mimas'' and ``Tethys'' would be impacted by their currently estimated mass of rock provided the initial mass of the transneptunian disk was $> \SI{20}{M_{\oplus}}$, but that ``Enceladus'' would be impacted by about a factor of 2 too little rock mass. We envision several possibilities to explain this discrepancy: 1) most of the mass was delivered by large impactors with radius $\sim \SI{100}{km}$ or larger \citep{charnoz09}. We estimate that stochastic variations could then produce the observed rock distribution in the system with a probability up to $10\%$. 2) Enceladus may have initially formed with more rock due to the presence of some large chunks in the rings, as suggested by \cite{charnoz11}. Or 3) the amount of rock colliding with each moon was different than the rock ultimately retained by each moon.

Per 3): impact velocities onto the spawned moons will be dominated by gravitational focusing by Saturn, and so will greatly exceed each moon's escape velocity. For a typical impact velocity $\sim 20$ km s$^{-1}$, the impactor will be destroyed and its rock shock heated to temperatures $\sim 2000$ to $8000$ K, implying a primarily melt-vapor state for all but the most highly oblique impacts \citep{pierazzo00}. While much of the ejecta in a hypervelocity cratering event, e.g., that originated from the target in the``far field'', has ejection velocities comparable to the target's escape velocity \citep{alvarellos05}, the impactor material itself will have an ejection velocity within a factor of several of the impact velocity \citep{melosh89,pierazzo00}. Thus most or all of the impacting material will initially escape the satellite, although most will still be in bound Saturn orbit \citep{movshovitz15}. For small ejecta sizes, as would be expected for droplets condensing from vapor, ejecta-ejecta collisions may occur on much smaller timescales than re-accretion onto the target, allowing ejecta to collisionally damp to a ring that overlaps the orbits of neighboring satellites. Thus while collisions with Saturn's inner satellites should effectively capture rock from external impactors into the inner satellite region, where the rock ultimately accretes will depend on the ejecta's size and velocity distribution and on its post-impact evolution. This will require follow-on models to assess.

External impactors would have also encountered the rings. Large impactors would pass through the rings, while those small enough to encounter a ring mass comparable to their own during a single passage could be directly captured.  For a ring at 700 Myr with a surface density $\sim $ few $\SI{e2}{g.cm^{-2}}$, this corresponds to impactor radii smaller than about $\SI{2}{\meter}$. The size distribution of small impactors during the late heavy bombardment is uncertain, but it is inferred to be quite shallow with a cumulative size index of $-1.5$, based on the cratering record on Iapetus \citep{charnoz09,movshovitz15}.  With  $0.002$ times the total impactor population expected in objects $\le \SI{1}{\kilo\metre}$ in radius \citep{movshovitz15}, the fraction in objects smaller than $\SI{2}{\meter}$ would be $\sim 10^{-7}$. This implies a relatively small total rock mass captured by the rings, comparable to or smaller than the upper limit on the rock in the rings today \citep{nicholson08}.

\section{Discussion}

We have simulated the spawning of inner moons at Saturn from a massive primoridal ice ring as it viscously evolves. Our model includes the viscous spreading of the rings driven by the effects of self-gravity, interaction between the rings and the satellites at Lindblad resonances, an explicit treatment of mutual interactions between the spawned moons including capture into mean motion resonance (MMR), and the evolution of spawned moon orbits due to tidal dissipation in Saturn.  For the latter we assume a tidal dissipation parameter for Saturn of $Q \sim 10^4$, consistent with traditional estimates \citep{murray99} but implying a slower rate of primordial orbital expansion than may apply to Saturn over recent decades \citep{lainey15}. We investigate the effects of the initial ring mass, the planet's early radius, the presence or absence of Dione and Rhea (which we assume formed separately from the rings) and the strength of tidal dissipation within the satellites.

We find that by $10^9$ years, the distribution of spawned moons masses and semi-major axes ressembles that of current Mimas, Enceladus and Tethys.  Spawned satellites grow initially by accreting moonlets just as they are spawned from the rings, and later, when their orbits have expanded away from the ring edge, by accreting larger objects, themselves the result of accretion of moonlets spawned from the rings.   We therefore observe a behavior comparable to the discrete and pyramidal regimes described in the analytical work of \cite{crida12}. We find that capture of inner spawned moons into MMR with outer moons acts to expand the orbits of the satellites beyond those expected if each object only interacts with its own strong resonances in the rings, as has been assumed in prior work \cite{charnoz11}.  Thus the outermost spawned moons in our simulations may reach distances comparable to that of Tethys in $10^9$ years even though we consider relatively slow tidal evolution.  

Inner moons spawned from an ice ring would initially contain little-to-no rock. Using the mass and semi-major axis distributions of spawned moons from our simulations, we estimate the mass of rock that would have been delivered to these inner moons during a late heavy bombardment (LHB) in the outer solar system. We find that external bombardment of the inner moons is expected to deliver a mass in rock comparable to the total rock in Mimas, Enceladus and Tethys today. In contrast, the same bombardment would have delivered a mass in rock to Dione and Rhea that is much smaller that the mass of rock in those moons today. The overall implication is that Saturn's inner moons were predominantly ice when they formed, while outer Dione and Rhea were already relatively rock-rich when they formed. We argue that this is most simply explained if the inner moons are a byproduct of a massive early ice ring, while outer Dione and Rhea formed separately, presumably from the circumplanetary disk that produced Titan.  

Some individual impacts during the late heavy bombardment may have been energetic enough to catastrophically disrupt Saturn's inner moons, with rapid re-accretion then likely \citep{movshovitz15}.  Thus the spawned moons in our simulations may be best viewed as the progenitors to Mimas, Enceladus and Tethys.  In our simulations that include pre-existing Dione and Rhea (Sets C and D), we find at $10^9$ years many resonant configurations (involving mostly 2:1 and 3:1 MMRs) between Dione or Rhea and inner satellites spawned from the rings. These configurations are found more frequently in cases that consider strong tidal dissipation in the satellites, because this tends to stabilize the resonant configurations. 

The integrations we perform are numerically intensive, and as such they involve several simplications.  The principal one is the simplicity of our Roche-interior disk model, which does not resolve its radial structure and assumes that it maintains a flat surface density profile at all time.  This is a much simpler treatment than the model of \cite{charnoz11}, but it allows us to explicitly model the dynamical evolution of the spawned satellites and their mutual interactions, which in particular allows for capture into MMRs, a feature absent from the \cite{charnoz11} model that proves important in our results. Further, we utilize an accelerated version of our code in simulating the system evolution from $10^8$ to $10^9$ yrs.

We find the total mass of external rock that collides with the inner spawned moons during the LHB to be compatible with the estimated \textit{total} mass of rock in the inner moons. Explaining the actual rock distribution in each of these three satellites (or their progenitors) is challenging, potentially requiring either a stochastic component of large impactors and/or some rocks in side the initial rings. In the case of the former, we estimate a few to $10\%$ likelihood of reproducing the observed rock distribution for impacts $\gtrsim \SI{100}{km}$ in radius \citep[e.g.][]{charnoz09}, in the limit that all the rock that collides with a moon is retained by the moon. This is likely a poor assumption for high-velocity impacts, and where the ejected material for each collision will ultimately be accreted should be considered by future work.
 
\newpage

\appendix
\section{Moment of inertia of a shell with variable density}{\label{AppendixMomentInertia}}

Saturn's current moment of inertia constant is estimated to be $K_{\saturn} \approx 0.23$ \citep{helled11,nettelmann13}. We can approximate this moment of inertia by considering that Saturn is a core surrounded by a gaseous shell with a density $\rho(r) = \rho_0(R_{\saturn}/r)^2$. The moment of inertia of the core of mass $M_{core}=\SI{20}{M_\oplus}$ and radius $R_{core}$ is $I_{core}=(2/5)M_{core}R_{core}^2$. We need to compute the moment of inertia of a shell with inner radius $R_1$ and outer radius $R_2$, and a density variation with distance $\hat{r}$ such that $\rho(\hat{r}) = \rho_0(R_2/\hat{r})^2$. In spherical coordinates, a point of the shell has coordinates $\hat{r}$, $\theta$ and $\phi$. Let $r$ denote the distance of the point to the axis of inertia. Then $\rho(r) = \rho_0(R_2\sin\theta/r)^2$. The moment of inertia is then computed using:

\begin{align}
I&=\int_{\hat{r}=R_1}^{R_2}\int_{\theta=0}^{\pi}\int_{\phi=0}^{2\pi}\rho\left(r\right)r^2\left(d\hat{r}\hat{r}d\theta\hat{r}\sin\theta d\phi\right)\\
&=\rho_0R_2^2\int_{\hat{r}=R_1}^{R_2}\hat{r}^2d\hat{r} \int_{\theta=0}^{\pi}\sin^3\theta d\theta \int_{\phi=0}^{2\pi}d\phi\\
&=\frac{8}{9}\pi\rho_0R_2^2\left(R_2^3-R_1^3\right)
\end{align}

The mass of the shell is:

\begin{align}
M_{shell}&=\int_{\hat{r}=R_1}^{R_2}\int_{\theta=0}^{\pi}\int_{\phi=0}^{2\pi}\rho\left(\hat{r}\right)\left(d\hat{r}\hat{r}d\theta\hat{r}\sin\theta d\phi\right)\\
&=\rho_0R_2^2\int_{\hat{r}=R_1}^{R_2}\hat{r} \int_{\theta=0}^{\pi}\sin\theta d\theta \int_{\phi=0}^{2\pi}d\phi\\
&=4\pi\rho_0R_2^2\left(R_2-R_1\right)
\end{align}

The moment of inertia of the shell can then be written as:

\begin{equation}
I_{shell}=\frac{2}{9}M_{shell}\left(R_{shell}^2+R_{core}R_{shell}+R_{core}^2\right).
\end{equation}

The moment of inertia constant of the whole planet is then:

\begin{equation}
K_p=\frac{I_{core}+I_{shell}}{M_{\saturn}R_{shell}^2}.
\end{equation} 

For current Saturn, using a core with density $\rho_{core} = \SI{10}{g.cm^{-3}}$ we get $K_p=0.234$, in very good agreement with the value quoted above. For our largest planet, $R_{shell}=\SI{1.5}{R_{\saturn}}$ and then $K_p=0.211$, close to the value for current Saturn. 

\newpage
\bibliography{references}

\end{document}